\documentclass[prd, 10pt,aps,nofootinbib, floatfix, notitlepage,twocolumn,longbibliography]{revtex4-2}

\usepackage{amsmath,amsfonts,amssymb,enumitem,fontawesome5}
\usepackage{mathrsfs}
\usepackage{multirow}
\usepackage{graphicx}
\usepackage[english]{babel} 
\usepackage{url} 
\usepackage{color,xcolor}
\usepackage{bbold}
\usepackage{booktabs}
\usepackage{adjustbox}
\usepackage[utf8]{inputenc} 
\usepackage[colorlinks=true,citecolor=blue,urlcolor=blue]{hyperref}
\usepackage[utf8]{inputenc} 
\usepackage[normalem]{ulem}

\renewcommand\sout[1]{\unskip}
\newcommand{\edit}[1]{\textcolor{black}{ #1}}

\begin{document}

\title{Quantifying Scalar Field Dynamics with DESI 2024 Y1 BAO Measurements}

\author{Kim~V.~Berghaus$^{1}$}

\author{Joshua~A.~Kable$^{2}$}

\author{Vivian~Miranda$^{2}$}

\affiliation{$^1$Walter Burke Institute for Theoretical Physics, California Institute of Technology, CA 91125, USA}
\affiliation{$^2$C.N. Yang Institute for Theoretical Physics, Stony Brook University, NY 11794, USA}

\begin{abstract}
Quintessence scalar fields are a natural candidate for evolving dark energy. Unlike the phenomenological $w_0w_a$ parameterization of the dark energy equation of state, they cannot accommodate the phantom regime of dark energy $w(z) < -1$, or crossings into the phantom regime. Recent baryon acoustic oscillation (BAO) measurements by the Dark Energy Spectroscopic Instrument (DESI) indicate a preference for evolving dark energy over a cosmological constant, ranging from $2.6\sigma-3.9\sigma$ when fitting to $w_0w_a$, and combining the DESI BAO measurements with other cosmological probes. 
In this work, we directly fit three simple scalar field models to the DESI BAO data, combined with cosmic microwave background anisotropy measurements and supernova data sets. 
We find the best fit \edit{model} to \sout{be} \edit{include a} $2-4\%$ \sout{of} kinetic scalar field energy $\Omega_{\rm scf,k}$, for a canonical scalar field with a quadratic or linear potential. \sout{with} \edit{However,} only the DESY-Y5 supernova data set combination \sout{showing} \edit{shows} a preference for \sout{dynamics} \edit{quintessence} over $\Lambda$CDM at the $95\%$ confidence level.
Fitting to the supernova data sets Pantheon, Pantheon+, DES-Y5, and Union3, we show that the mild tension ($n_{\sigma}< 3.4 $) under $\Lambda$CDM emerges from a BAO preference for smaller values of fractional mass-energy density $\Omega_m < 0.29$,  while all supernova data sets, except for Pantheon,  prefer larger values, $\Omega_m > 0.3$. The tension under $\Lambda$CDM remains noticeable ($n_{\sigma} <2.8$), when replacing two of the DESI BAO redshift bins with effective redshifts $z_{\text{eff}} =0.51$, and $z_{\text{eff}}= 0.706$ with comparable BOSS DR 12 BAO measurements at $z_{\text{eff}} =0.51$, and $z_{\text{eff}}= 0.61$. Canonical scalar fields as dark energy are successful in mitigating that tension.  
\end{abstract}

\maketitle

\date{\today}

%\tableofcontents

% --------------------------------------------------------------------
% --------------------------------------------------------------------
% --------------------------------------------------------------------
\section{Introduction}
\label{Sec:Introduction}
% --------------------------------------------------------------------
% --------------------------------------------------------------------
% --------------------------------------------------------------------

Recent measurements of baryon acoustic oscillations (BAO) in galaxy, quasar, and Lyman-$\alpha$ forest tracers from the first year of observations
from the Dark Energy Spectroscopic Instrument (DESI) find hints towards an evolving equation of state of dark energy \cite{DESI:2024mwx}. Spanning a redshift range from $0.1 < z < 4.16$, the DESI BAO measurements provide the first measurements of the transverse comoving distance for redshifts larger than $z > 2.3$. DESI BAO data alone are consistent 
with the concordance $\Lambda$CDM model of cosmology in which dark energy is described by a cosmological constant such that the equation of state of dark energy is $w = -1$. However, when combined with supernova data, a mild tension emerges under $\Lambda$CDM.

Allowing for a time-varying dark energy equation of state, parameterized by $w(z) = w_0 + \left(1-1/(1+z)\right) w_a $, combinations of DESI with cosmic microwave background (CMB) or type Ia supernovae measurements find a preference for $w_0 > -1$, and $w_a < 0$, ranging from $2.6\sigma - 3.9\sigma$, depending on the data combination and choice of supernova dataset \cite{DESI:2024mwx}. This preference is consistent with results obtained previously by both the Union3 \cite{Rubin/etal:2023} and the Dark Energy Survey (DES) Y5 \cite{DES/SNY5} supernova compilations when combined with \textsl{Planck} CMB primary anisotropy \cite{Planck/results} and BAO data from BOSS \cite{Alam/etal:2017} and eBOSS \cite{alam/etal:2021}, which suggests this preference is not likely to be associated with a systematic in solely the DESI data. With the DESI year 1 data release \cite{DESI:2024mwx,DESI:2024uvr,DESI:2024lzq} providing the first influx of precision data from Stage IV \cite{Albrecht:2006um} dark energy experiments such as DESI, Euclid \cite{2016SPIE.9904E..0OR},  and the Vera Rubin Observatory \cite{LSST:2008ijt} (previously referred to as Large Synoptic Survey Telescope (LSST)), this first potential hint of dynamical dark energy is intriguing. 

From a theoretical perspective, the small value of the cosmological constant $\Lambda$, in the $\Lambda$CDM model of cosmology suffers fine-tuning \cite{Weinberg:1988cp}; that is quantum corrections of $\Lambda$ are larger than the observed value. The dynamical evolution of dark energy holds a promise to alleviate that tuning \cite{Abbott:1984qf, PhysRevLett.52.1461, Alberte:2016izw, PhysRevD.65.126003, Graham:2017hfr, Graham:2019bfu}, motivating dynamical dark energy with a redshift-dependent equation of state. Swampland conjectures \cite{Vafa:2005ui,Ooguri:2006in, Ooguri:2018wrx, Bedroya:2019snp}, a set of conjectured criteria for theories in the quantum gravity landscape, also favor dynamical dark energy over a cosmological constant as a fundamental description of dark energy. 

Scalar fields are a fundamental description of evolving dark energy that connects the property of a dominant energy density with negative pressure to a particle description. 
Quintessence scalar fields with canonical kinetic terms are theoretically well-motivated and exhibit thawing or tracking behavior \cite{peebles,Zlatev:1998tr,Wetterich:1994bg,Wang:1999fa,Park:2021jmi}. In this paper, we focus on scalar fields that exhibit thawing behavior; that is, the Hubble friction severely overdamps the scalar field at early times such that the field is effectively frozen with an equation of state approximately equal to negative one. At smaller redshifts,  Hubble friction becomes less efficient, and the scalar field's kinetic energy grows, leading to an equation of state that increases over time such that $w(0) > -1$. Simple renormalizable power-law potentials exhibit this thawing dynamic.

On the other hand, exponential scalar field potentials, for example, have tracking behavior, i.e., the dark energy density tracks the dominant energy density of the universe \cite{PhysRevD.37.3406}, and the equation of state is larger than negative one at early times $w(z)> -1$, and approaches $w(0) \approx -1 $ today. 
A third model category, phantom scalar fields with non-canonical kinetic terms, exhibit an evolving equation of state that is less than negative one. This behavior requires a negative kinetic energy term that can be thought of as a scalar field rolling up its potential. Phantom models have theoretical pathologies \cite{Caldwell:1999ew,PhysRevLett.91.071301,Caldwell:2003vq, Nojiri:2005sx}, one of which is a violation of the null energy condition, that leads to the future death of the universe in a big rip \cite{Caldwell:2003vq}. 

In light of the DESI BAO measurements preference for $w_0 > -1$, 
we fit the BAO measurements to canonical scalar fields with thawing behavior. 
This well-motivated description of dynamical dark energy provides a hypothesis for the redshift dependence of the dark energy equation of state, which is parameterized by one parameter beyond $\Lambda$CDM, the value of $w(0)$, which directly maps onto the kinetic energy of the scalar field today, $\Omega_{\text{scf,k}}$. Unlike the $w_0w_a$-parameterization used in the DESI analysis, thawing scalar field models asymptotically approach an equation of state of negative one at large redshift, such that the phantom-regime, $w(z) < - 1$ is not part of the possible dynamics. By doing so, they impose an arguably well-motivated theory prior, which cuts out the dark energy phantom-regime and phantom crossings. In our analysis, we fit the simplest canonical scalar fields with a linear or quadratic potential to the DESI BAO data, directly addressing the points raised in \cite{Cortes:2024lgw}, by quantifying the evidence for evolving dark energy in a physical model beyond $w_0w_a$. 

Notably, taking as an example the best fit to $w_0w_a$ with DESI BAO data + CMB + Union3, $w_0 = -0.64 \pm 0.11; \, w_a = -1.27^{+0.40}_{-0.34}$, indicates $w_0 > -1$ at $3.5\sigma$, which points towards a possible preference of thawing scalar field behavior. However, the best-fit crosses over into the phantom regime, $w(z) < -1$, for redshifts exceeding $z = 0.4$, which is not possible for canonical scalar field dark energy.  
We analyze the preference of DESI BAO data in combination with CMB and supernova datasets for thawing scalar fields over $\Lambda$CDM, which we quantify by evaluating the preference for scalar field dynamics, e.g., nonzero kinetic scalar field energy or nonzero dark energy radiation \cite{Berghaus:2020ekh,Berghaus:2023ypi}. Quintessence has already been invoked as an explanation of the DESI results \cite{Tada:2024znt, Yin:2024hba}. Here, we perform a complete fit for concrete scalar field models to the DESI BAO data, \edit{whereas previous works have projected select results of the $w_0w_a$ fit performed by the DESI collaboration onto quintessence cosmologies.  }

We focus on canonical scalar fields with a quadratic or linear potential, as well as a scenario in which the dynamical component is comprised of dark radiation sourced by the scalar field, dark energy radiation \cite{Berghaus:2020ekh,Berghaus:2023ypi}. 

This paper is organized as follows. In Sec.~\ref{sec:background}, we briefly review the quantities measured by BAO and supernova light curve measurements and discuss how they characterize the expansion history of our universe and, consequently, the equation of state of dark energy. In Sec.~\ref{sec:models}, we discuss the three scalar field cosmologies we fit to: quadratic (SCF QUAD), linear (SCF LIN), and dark energy radiation (SCF DER). We derive how the scalar field dynamics map onto the usual cosmological parameterizations of the dark energy equation of state. In Sec.~\ref{sec:data}, we describe our methodology and dataset combinations, which include CMB measurements, two combinations of BAO measurements, as well as four supernova datasets. 
In Sec.~\ref{sec:results}, we present our results, the marginalized posteriors for the three scalar field cosmologies, as well as the mean values of the preferred scalar field dynamics and their associated uncertainties, $\Omega_{\text{scf,k}}$ or $\Omega_{\text{der}}$. We find that evidence for evolving dark energy meets the $95\%$ confidence level for a simple canonical scalar field with a quadratic or linear potential only in combination with the DES-Y5 supernova data set. Constraining the sum of neutrino masses under the three scalar field cosmologies, we find that the bounds become stronger, but remain comparable to those derived under $\Lambda$CDM. 
We also quantify the mild tension between BAO and supernova measurements under $\Lambda$CDM, and the reduction of that tension under the scalar field cosmologies, as well as $w_0w_a$.
In Sec.~\ref{sec:concl}, we summarize our main findings and conclude with canonical scalar fields being promising candidates for explaining the DESI BAO measurements.

In Appendix \ref{sec:w0wa}, we show results for various additional $w_0w_a$ parameterizations with priors chosen that mimic thawing, tracking, and phantom scalar field behavior, finding that thawing behavior is preferred over tracking and phantom dynamics.
We also perform a principle component analysis (PC), quantifying the difference two of the DESI redshift bins make in the mild tension of BAO and supernova measurements under $\Lambda$CDM. 

% --------------------------------------------------------------------
% --------------------------------------------------------------------
% --------------------------------------------------------------------
\section{Background}
\label{sec:background}
% --------------------------------------------------------------------
% --------------------------------------------------------------------
% --------------------------------------------------------------------

The baryon acoustic oscillation pattern, formed from pressure waves in the baryon-photon fluid prior to the decoupling of the photons from the baryons around $z \simeq 1100$, is imprinted in both the distributions of the CMB photons and the matter in the universe. We refer to the measurements of the baryon acoustic oscillation pattern observed in visible matter as the BAO. Both the CMB and BAO measurements are sensitive to the size of the sound horizon at the time of decoupling, $r_d$,\footnote{The BAO are sensitive to the size of the sound horizon when the baryons decouple from the photons, $z_d \simeq 1060$. Because there were many orders of magnitude more photons than baryons, a relatively small amount of photons was sufficient for the baryons to stay coupled, delaying when the baryon acoustic oscillation pattern was set in the distribution of baryonic matter. This is referred to as the drag epoch. } and the expansion of the universe. The sound horizon is determined by 

\begin{equation}
r_d = \int_{z_d}^{\infty} \frac{c_s(z)}{H(z)}dz,
\end{equation}
where $c_s$ is the speed of sound prior to recombination, determined by the ratio of baryons to photons. 
\begin{equation}
c_s(z) = \frac{c}{\sqrt{3 \Big(1+ \frac{3\rho_B(z)}{4\rho_{\gamma}(z)} \Big)}} \, .
\end{equation}

More specifically, the BAO measures a combination of the transverse comoving distance and the sound horizon, $D_M(z)/r_d$, as well as the equivalent distance variable and the sound horizon, $D_H(z)/r_d$. The transverse comoving distance and the equivalent distance variable are respectively defined as
\begin{equation}
D_M(z) = \frac{c}{H_0\sqrt{\Omega_K}} \sinh \left( \sqrt{\Omega_K} \int_0^z \frac{dz'}{H(z)/H_0} \right) \, ,
\end{equation}
and 
\begin{equation}
D_H(z) = \frac{c}{H(z)} \, ,
\end{equation}
\begin{equation}
\label{eq:Hofz}
H(z) = H_0 \sqrt{\Omega_m(1+z)^3 + \Omega_{\rm{DE}}^{3(1+w(z))} + \Omega_K (1+z)^2 }  \, 
\end{equation}
where $\Omega_K$ accounts for nonzero curvature, and we have assumed a general dark energy, $\Omega_{\rm{DE}}$ with equation of state parameter that in general may vary with redshift, $w(z)$. For conceptual clarity, in Eq.~\eqref{eq:Hofz} we have neglected the negligible amount of radiation around at redshifts measured by BAO, as well as the effects that massive neutrinos have on the transfer of energy densities between radiation and matter. 

Combining measurements of the CMB and the BAO is a powerful probe of cosmology because both measure the same sound horizon scale at different points in the evolution of the universe. The CMB anisotropy measures the distribution set at $z\simeq 1100$, while DESI BAO, for example, covers the range $ 0.4\leq z \leq 4.16$. This has the effect of breaking degeneracies between parameters and thus providing tighter parameter constraints. 

Another important cosmological probe is the measurement of the luminosity distances, $d_L(z)$, of type Ia supernova as a function of redshift. These measurements are sensitive to the expansion history of the universe by measuring the flux of light, $F$, from an object with known luminosity $L$. In particular, 
\begin{equation}
F  = \frac{L}{4\pi D_M(z)^2(1+z)^2} = \frac{L}{4\pi d_L(z)^2} \, .
\end{equation}
The two factors of $(1+z)$ arise because the expansion of the universe causes light to redshift, losing energy, as well as changes to the rate of reception of photons. In the latter case, two photons emitted at times separated by $\Delta t_e$, will be observed to be separated in time by $\Delta t_o = (1+z)\Delta t_e$.
 Hence, the luminosity distance is \cite{Scolnic/etal:2018}
\begin{equation}
d_L(z) = \frac{c (1+ z)}{H_0\sqrt{\Omega_K}} \sinh \left( \sqrt{\Omega_K} \int_0^z \frac{dz'}{H(z)/H_0} \right) \, .
\end{equation} 
Including CMB anisotropy measurements breaks the degeneracy between $H_0$ and $r_d$. They also constrain the baryon density, as well as $\Omega_m$.  

The flat $\Lambda$CDM model sets $\Omega_K = 0$, which implies $\Omega_{\rm{DE}} = \Omega_{\Lambda} = 1- \Omega_m$, as well as $w = -1$ for the dark energy equation of state. In the $w_0w_a$-parameterization the equation of state is given by $w(z) = w_0 +(1-\frac{1}{1+z}) w_a$, which allows the dark energy density to evolve in redshift. In this work, we go beyond phenomenological parameterizations of the equation of state of dark energy and fit BAO and supernova directly to scalar field models.

% --------------------------------------------------------------------
% --------------------------------------------------------------------
% --------------------------------------------------------------------
\section{Scalar Field Models}
\label{sec:models}
% --------------------------------------------------------------------
% --------------------------------------------------------------------
% --------------------------------------------------------------------

\subsection{Quintessence}
\label{subsec:scalar_fields}
We consider a canonical scalar field $\phi$ to comprise the dark energy density whose evolution is governed by its scalar field potential $V(\phi)$ and the Hubble expansion rate of the universe, $H = \dot{a}/a$, where $a$ is the scale factor, and $\dot{a}\equiv \frac{da}{dt}$. We take the entire energy density of dark energy to be comprised of the scalar field such that there is no additional cosmological constant. We also assume that there is no curvature, $\Omega_K = 0$. 

The scalar field's homogeneous equation of motion is 
\begin{equation}
    \ddot{\phi}(t) +3 H(t)  \dot{\phi}(t) +\frac{dV}{d\phi} = 0      \, . 
\end{equation}
The Hubble expansion under this hypothesis is given by 
\begin{equation}
H(z) = H_0 \sqrt{\Omega_m(1+z)^3 + \frac{V(\phi(z)) +\dot{\phi}(z)^2/2 }{\rho_c}} \, , 
\end{equation}
where
\begin{equation}
\Omega_m + \frac{V(\phi(0)) + \dot{\phi}(0)^2/2}{\rho_c} = 1.
\end{equation}
The critical density of the universe today is defined as $\rho_c \equiv 3 M_{pl}^2 H^2_0$.
For two choices of scalar field potential, we quantify the preference for non-zero kinetic energy that the DESI BAO data exhibits by itself and in combination with other datasets. We define
\begin{equation}
\Omega_{\text{scf,k}} \equiv \frac{\dot{\phi}(z =0)^2/2}{\rho_c} \, ,
\end{equation}
as the fraction of kinetic scalar field energy today over the critical density of the universe today. \edit{The kinetic scalar energy at early times is fixed by attractor initial conditions that effectively freeze the field due to large Hubble friction such that $\dot{\phi}(z\to \infty) \to 0$ for the linear and quadratic potentials we consider. Thus, $\Omega_{\text{scf,k}}$ is a derived parameter, which is fully determined by the slope of the potential $V$.}

\begin{figure}[!tbp]
\centering
\raisebox{-0.5\height}{\includegraphics[width=0.485\textwidth,angle=0]{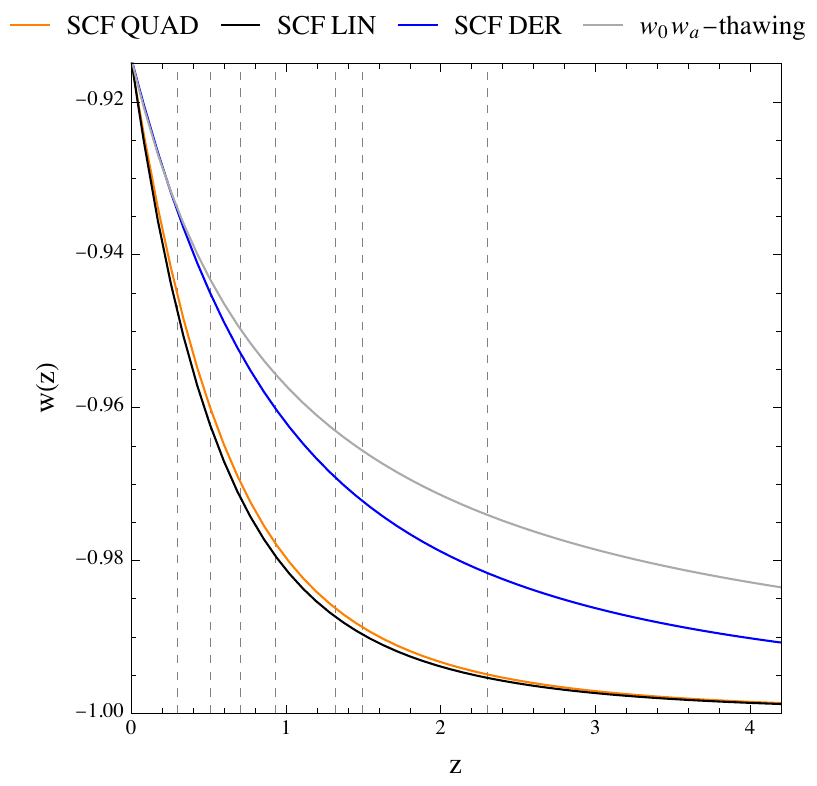}}
\vspace{-0.325cm}
\hfill
\caption{Comparison of the dark energy equation of state $w(z)$ as a function of redshift for the three scalar field cosmologies described in Sec.~\ref{sec:models} for the same value $w(0)=-0.914$, as well as the phenomenological $w_0w_a$ thawing parameterization, described in Appendix \ref{sec:w0wa}, that mimics thawing scalar field dynamics. The curves correspond to values of $\Omega_{\text{scf,k}} = 0.03$, $\Omega_{\text{der}} = 0.09$, and $w_{\text{thawing}} = -0.914$, for fixed $\Omega_m= 0.3$ for all curves. The dashed vertical lines indicate the effective redshifts $z_{\text{eff}}$ of the DESI BAO measurements.}
\label{fig:wz}
\end{figure}

The equation of state of the scalar field dark energy as a function of redshift is determined by
\begin{equation}
    w(z) \equiv \frac{p(z)}{\rho(z)}
    = \frac{\frac{ {\dot{\phi}(z)}^2}{2} 
    -V(\phi(z))}{\frac{ {\dot{\phi}(z)}^2}{2} 
    + V(\phi(z))}  \, . 
\end{equation}
Analytical expressions for $w(z)$, valid in the regime in which $w(z)$ is near $-1$, can be found in \cite{Scherrer:2007pu,Dutta:2009yb}.
In that regime in which $\dot{\phi^2}/2 \ll V$, the equation of state simplifies to 
\begin{equation}
w(z) \approx -1 + 2\Delta w(z) = -1 + \frac{\dot{\phi}(z)^2}{V(\phi(z))} \, .
\end{equation}
Comparing this with the $w_0w_a$ parameterization 
\begin{equation}
w(z) = w_0 + \left(1-\frac{1}{1+z}\right) w_a \, ,
\end{equation}
one can identify 
\begin{equation}
w_0 + 1 \approx \frac{\dot{\phi}(0)^2}{V(\phi(0))} \, ,
\end{equation}
which indicates that the preference for a positive deviation from $w_0 = -1$ found in the DESI BAO data may map onto a preference for non-zero kinetic scalar field energy today at redshift $z = 0$. 
Translating the scalar field description in a flat universe onto the $w_0w_a$ parameterization, one finds
\begin{equation}
w_0 \approx  \frac{2\Omega_{\text{scf,k}}}{1- \Omega_m} -1 \, , \text{and} 
\end{equation}
\begin{equation}
w_a = 1 +w_0 \, ,
\end{equation}
where we have approximated $V(\phi(0))/\rho_c \approx 1 - \Omega_m$. 
Many scalar field potentials have been considered historically in the context of quintessence dark energy (see, for instance, \cite{peebles,Sahni:1999qe,Barreiro,Ferreira, Caldwell:2005tm}). In this work, we focus on two simple scenarios that describe the local shape of the potential. For this purpose, we consider a quadratic scalar field potential
\begin{equation}
V(\phi) = \frac{1}{2} m^2 \phi^2 \,,\,\,\,\,\,\,\,\,\,\, (\text{SCF QUAD}) 
\end{equation}
 and a linear potential scalar field potential 
\begin{equation}
\label{eq:lin}
V(\phi) = - C \phi \, , \,\,\,\,\,\,\,\,\,\text{(SCF LIN)} \end{equation}
which we label as  `SCF QUAD', and `SCF LIN', henceforth. Both choices of potential exhibit thawing behavior, implying that the kinetic energy of the scalar field is largest today, and the dark energy equation of state asymptotes to negative one at large redshift as shown in Fig.~\ref{fig:wz}. The scalar field dynamics cannot accommodate the best fit for $w_0$ and $w_a$, found with the DESI BAO dataset in combination with supernova data due to the preference for a crossing into the phantom regime $w(z) < -1 $ which is not possible in a standard canonical scalar field model. We vary the linear slope $C \in [2.2 \times 10^{-8}, 2.4 \times 10^{-7}]$ of the scalar potential in units of $[M_{\text{pl}}\text{Mpc}^{-2}]$. The steepest slope in this prior allows for up to $\Omega_{\text{scf,k}} \approx 0.2$, and the shallowest slope allows for values as low as $\Omega_{\text{scf,k}} = 0.002$. We do not expand to smaller priors due to numerical instabilities in our modified CLASS implementation. 
For the quadratic potential we vary $m \in [1\times 10^{-5}, 1 \times 10^{-3}]$ in units of $[\text{Mpc}^{-1}]$. The largest mass allows for up to $\Omega_{\text{scf,k}} \approx 0.6$, and the smallest value of the prior corresponds to $\Omega_{\text{scf,k}} \approx 0.0001$ \edit{($w_0 = -0.9997$)}, effectively asymptoting to $\Lambda$CDM, \edit{which corresponds to $\Omega_{\text{scf,k}} =0 $ ($w_0 = -1$). }

The masses that feature evolving dark energy with $\mathcal{O}(1)$ dynamics are $m \sim H_0 \approx 10^{-33} \, \text{eV}$.

% --------------------------------------------------------------------
% --------------------------------------------------------------------
% --------------------------------------------------------------------
% --------------------------------------------------------------------
% --------------------------------------------------------------------
% --------------------------------------------------------------------

% --------------------------------------------------------------------
% --------------------------------------------------------------------
% --------------------------------------------------------------------
\begin{figure*}
\centering
\begin{tabular}{c}    
    \begin{tabular}{@{}c@{}}
        \includegraphics[width=0.485\textwidth,angle=0]{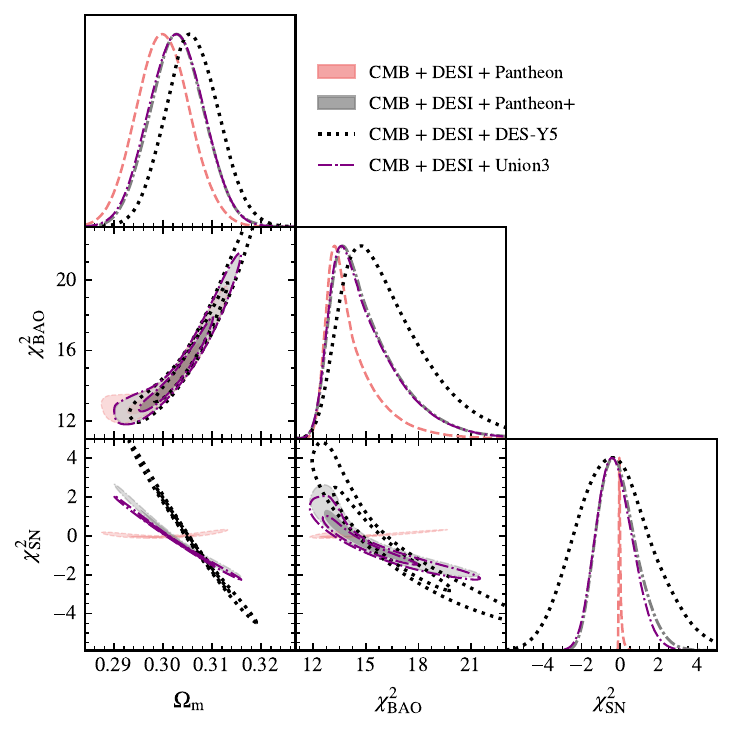}
        \includegraphics[width=0.485\textwidth,angle=0]{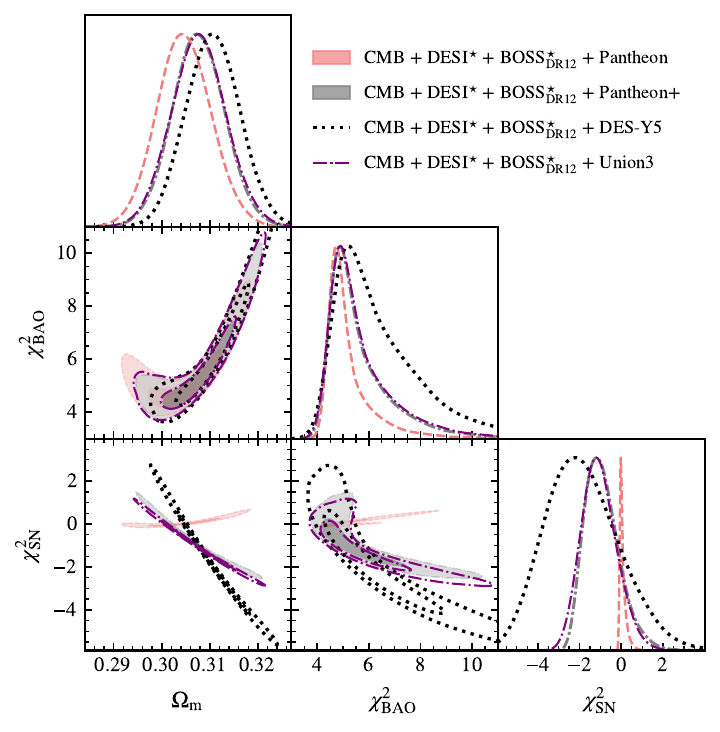} 
    \end{tabular} \\
 \includegraphics[width=0.8\textwidth,angle=0]{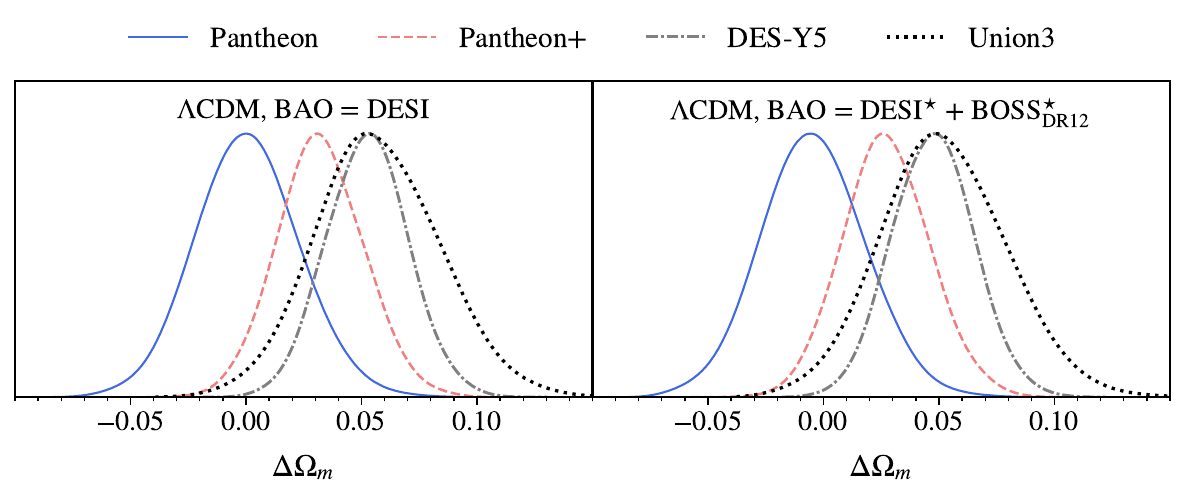}
\end{tabular}
\vspace{-0.325cm}
\hfil
\caption{(Top Panels) Supernova and BAO $\chi^2$-posteriors for $\Lambda$CDM in combination with CMB Planck and various supernova data sets. Two BAO data set combinations are shown, DESI to the left, and DESI$^{\star}$+BOSS$^{\star}_{\rm DR12}$, where we have replaced the DESI BAO $z_{\text{eff}} =0.51$, and $z_{\text{eff}} =0.706$ data points with comparable redshift BOSS DR12 measurements. The CMB Planck data has a reduced $\ell < 1296$ range to avoid nonlinear lensing effects. The opposing preference for $\Omega_m$ in $\chi^2_{\text{SN}}$ and $\chi^2_{\text{BAO}}$ indicates a mild tension of the datasets under $\Lambda$CDM. The shown supernova $\chi^2_{\text{SN}}$ for Pantheon, Pantheon+, DES-Y5, and Union3 have, respectively, the constant offset $\chi^2_{\rm SN, offset} = (1034.8, 1405.7, 1648.4, 31.5)$. (Bottom Panels) Probability distribution of the difference $\Delta \Omega_m$ between $\Omega_m$ density derived from CMB + BAO versus SN-only MCMC chains. Replacing DESI bins at $z_{\text{eff}} =0.51$, and $z_{\text{eff}} =0.706$ with BOSS DR12 measurement does not significantly reduces the tension on $\Omega_m$ values between CMB+BAO and newer Type-Ia supernova measurements that drive the detection of dynamical dark energy. Appendix \ref{sec:w0wa} shows that the CMB+DESI combination is perfectly compatible with the cosmological constant as long as type-Ia supernova constrains the cold dark matter density to be $\Omega_m \approx 0.289$, which is the range preferred by the Pantheon data set.
}
\label{fig:LCDM}
\end{figure*}
% --------------------------------------------------------------------
% --------------------------------------------------------------------
% --------------------------------------------------------------------

% --------------------------------------------------------------------
% --------------------------------------------------------------------
% --------------------------------------------------------------------
\subsection{Dark Energy Radiation}
\label{subsec:der}
% --------------------------------------------------------------------
% --------------------------------------------------------------------
% --------------------------------------------------------------------

Dark Energy radiation is a novel description of dark energy, which proposes that the dynamical component of dark energy is dominated by a thermal bath of relativistic particles sourced by thermal friction from a slowly rolling scalar field \cite{Berghaus:2020ekh, Berghaus:2023ypi}.
Such a model is motivated by considering couplings of the dark energy scalar field to other light particles \cite{Berera:1999ws, Bastero-Gil:2016qru, Berghaus:2019whh}.
For example, an axion-like field coupling to non-Abelian gauge fields leads to the phenomenology of dark energy radiation \cite{Berghaus:2020ekh, Berghaus:2023ypi,Berghaus:2019cls, Berghaus:2019whh}. The homogeneous equations governing the coupled evolution of the scalar field and the dark radiation are 
\begin{equation}
    \ddot{\phi}(t) +(3 H(t) +\Upsilon)  \dot{\phi}(t) +\frac{dV}{d\phi} = 0      \, , \nonumber
\end{equation}
\begin{equation}
    \dot{\rho}_{\text{der}}(t) +4H(t) \rho_{\text{der}}(t) = \Upsilon {\dot{\phi}(t)}^2\,,      \,\,\,\, \text{(SCF DER)} 
\end{equation}
We follow a toy model approach of dark energy radiation that treats the thermal friction coefficient $\Upsilon$ as a constant. We also choose our priors such that $\Upsilon \gg H$. In this regime, the kinetic energy of the scalar field is highly suppressed, $\dot{\phi}^2/2 \ll \omega_{\text{der}}$, and there is a degeneracy between the choice of scalar field potential, and friction coefficient, e.g., a steeper slope with larger friction $\Upsilon$ leads to the same phenomenology as a shallower slope with smaller friction coefficient. 
The Hubble parameter is given by 
\begin{equation}
H(z) = H_0 \sqrt{\Omega_m(1+z)^3 + \frac{V(\phi(z)) +\rho_{\text{der}}(z) }{\rho_c}} \, , 
\end{equation}
where 
\begin{equation}
\Omega_m + \frac{V(\phi(0)) + \rho_{\text{der}}(0)}{\rho_c} = 1.
\end{equation}

One extra parameter beyond $\Lambda$CDM, the amount of dark energy radiation today, defined as 
\begin{equation}
\Omega_{\text{der}} \equiv \frac{\rho_{\text{der}}(z=0)}{\rho_c} \, ,
\end{equation}
parameterizes this model. The behavior of dark energy is comprised of the sum of the scalar field and the dark energy radiation $\rho_{\text{der}}$. The equation of state is given by
\begin{equation}
w(z) \equiv \frac{p_\phi(z)+p_{\text{der}}(z)}{\rho_{\phi}(z) +\rho_{\text{der}}(z) } \, ,
\end{equation}
which due to $\dot{\phi}^2/2 \ll \rho_{\text{der}}$, simplifies to 
\begin{equation}
w(z) \approx -1 +2\Delta w(z) = -1 + \frac{1}{3}\frac{\rho_{\text{der}}(z)}{V(\phi(z))} \, .
\end{equation}
A deviation in the $w_0$ parameter from negative one indicates a nonzero component of dark energy radiation  
\begin{equation}
w_0 \approx \frac{1}{3} \frac{2 \, \Omega_{\text{der}}}{1-\Omega_m} -1  \, .
\end{equation}
Compared to quintessence scalar field models in which kinetic energy comprises the dynamical component described in Sec.~\ref{subsec:scalar_fields}, the equation of state of dark energy in dark energy radiation asymptotes slower to negative one at larger redshifts. This behavior is illustrated in Fig.~\ref{fig:wz}.
The dynamical evolution of dark energy radiation is insensitive to the choice of scalar field potential, due to the additional thermal friction severely overdamping the scalar field. Consequently, we explore the phenomenology of dark energy radiation on the example of a single choice for the scalar field, which we take to be linear, as given by Eq.~\eqref{eq:lin}. Due to the degeneracy between slope and friction, we fix the linear slope to be $C = 1\times 10^{-6}M_{\text{pl}}\text{Mpc}^{-2}$, and vary the friction coefficient $\Upsilon$ in units of [$\text{Mpc}^{-1}$] as $\log_{10}\Upsilon \in [-1,1]$. The lower prior limit of $\Upsilon$ allows for $\Omega_{\text{der}} \approx 0.1$, and the upper limit corresponds to the asymptotic limit of a frozen scalar field with $\Omega_{\text{der}} \approx 0$. We refer to this model as SCF DER.    

% --------------------------------------------------------------------
% --------------------------------------------------------------------
% --------------------------------------------------------------------
\begin{figure*}[!tbp]
\centering
\raisebox{-0.5\height}{\includegraphics[width=0.485\textwidth,angle=0]{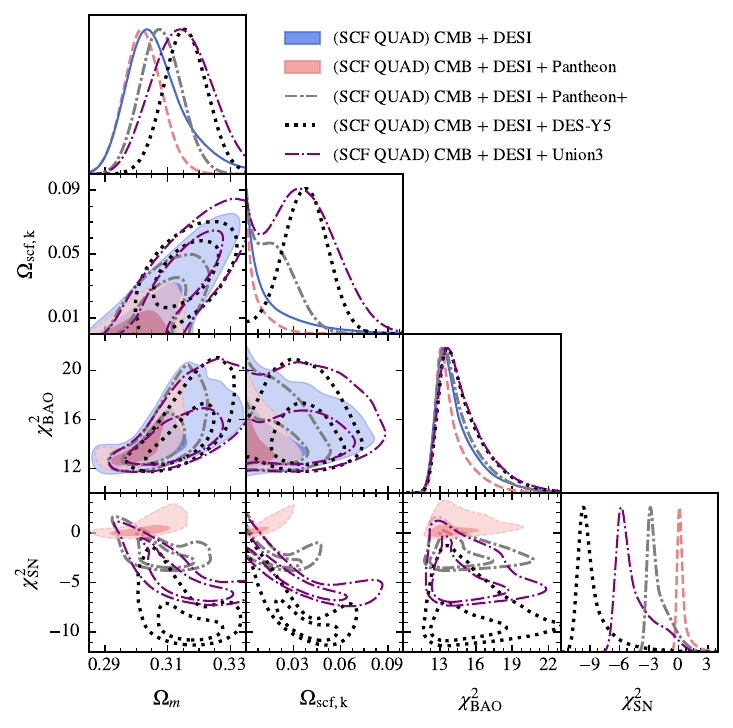}}
\raisebox{-0.5\height}{\includegraphics[width=0.485\textwidth,angle=0]{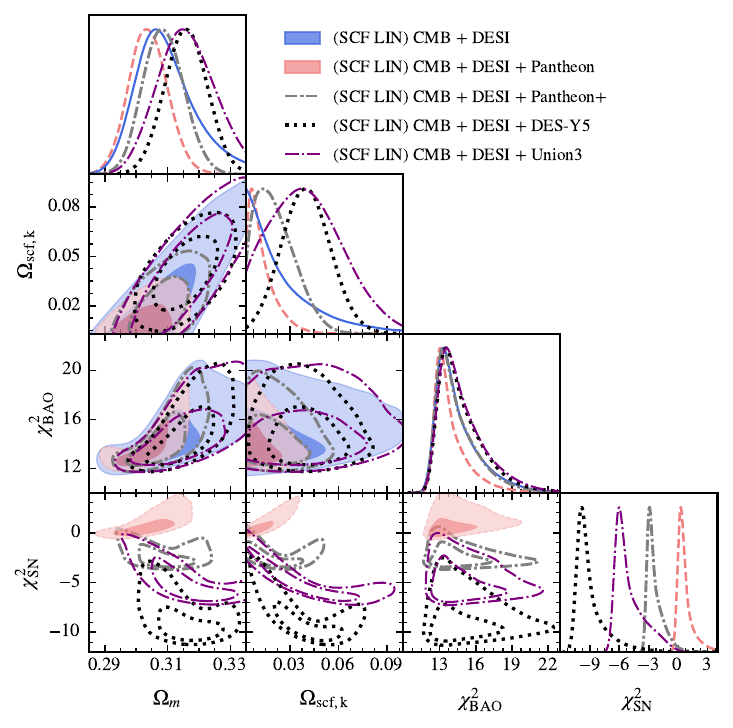}}
\vspace{-0.325cm}
\hfill
\caption{Marginalized posteriors for a scalar field with a quadratic potential (SCF QUAD) to the left and a linear potential (SCF LIN) to the right. The supernova $\chi^2_{\text{SN}}$ are shown with the same offset as before. The models allow a better fit to supernova data for smaller $\Omega_m$, easing the tension under $\Lambda$CDM shown in Fig.~\ref{fig:LCDM}. %For SCF QUAD, the $95\%$ confidence interval does not include the $\Lambda$CDM limit ($\Omega_{\text{scf,k}} = 0$) for any of the dataset combinations. 
%For SCF LIN cases with supernova data, the $95\%$ confidence interval only includes the $\Lambda$CDM limit when Pantheon data are used.
}  
\label{fig:SCF}
\end{figure*}
% --------------------------------------------------------------------
% --------------------------------------------------------------------
% --------------------------------------------------------------------

% --------------------------------------------------------------------
% --------------------------------------------------------------------
% --------------------------------------------------------------------
\section{Data and Methodology}
\label{sec:data}
% --------------------------------------------------------------------
% --------------------------------------------------------------------
% --------------------------------------------------------------------

We compare the scalar field dark energy cosmologies outlined in Sec.~\ref{sec:models} (SCF QUAD, SCF LIN, SCF DER) as well as the phenomenological $w_0w_a$ model to a standard flat $\Lambda$CDM universe. In Appendix \ref{sec:w0wa}, we also examine more phenomenological dark energy parameterizations.
In all cases, we vary the usual six $\Lambda$CDM base parameters: 
the amplitude $A_s$ of the primordial curvature power spectrum at $k=0.05 $Mpc$^{-1}$ as $10^9A_s$; 
the tilt $n_s$ of this spectrum; 
the angular size $\theta_*$ of the sound horizon, %($\theta_{\rm MC}$ for CAMB runs); 
the physical density $\Omega_bh^2$ of baryons; 
the physical density $\Omega_ch^2$ of cold dark matter;
and the optical depth $\tau$ to reionization, where we have chosen broad, uninformative priors.
We adopt the standard neutrino description with one massive ($m_{\nu} = 0.06$eV) and two massless neutrinos where we quantify the constraints on the sum of the neutrino masses under our scalar field dark energy cosmologies. In some select cases, we study varying the sum of the neutrino masses. 

We modified the Boltzmann code CLASS \cite{class2} to include the scalar field dark energy cosmologies described in Sec.~\ref{sec:models}. Our modified version of CLASS is publicly available\footnote{
\href{https://github.com/KBerghaus/class_der}{\tt https://github.com/KBerghaus/class\_{der}}  }. For all $\Lambda$CDM and all $w_0w_a$ models explored in this work, we use the \texttt{CAMB} Boltzmann code \cite{Lewis_2000}. To determine the posterior distributions for the various model parameters, we perform Markov Chain Monte Carlo (MCMC) runs using the publicly available \texttt{Cobaya} code \cite{Cobaya/paper,Torrado:2020dgo}. To assess the convergence of the MCMC chains, we use a Gelman-Rubin converge criterion of $R-1 = 0.02$. 

To constrain cosmological models, we primarily use observational data from CMB, BAO, and supernova. For CMB, we use the \textsl{Planck} 2018 multifrequency half-mission TT, TE, and EE power spectra \cite{Planck/overview, Planck/spectra, Planck/results}. For each of the TT, TE, and EE power spectra, we make a multipole cut so that we only include $\ell \leq 1296$. We do this to avoid nonlinear lensing effects as we do not have the modeling of nonlinear scales for the scalar field dark energy models. In addition, we use \textsl{Planck} $\ell < 30$ TT and EE data.

For BAO, we primarily use the DESI DR1 BAO results \cite{DESI:2024mwx}\footnote{Note that the effective redshifts in Table 1. of \cite{DESI:2024mwx} are rounded and introduce a significant truncation ($\Delta \chi^2 \sim 2$) error when used directly instead of the official DESI likelihood, which was released within the \textsc{cobaya} package only after \cite{DESI:2024mwx}.}, which include measurements of the BAO signal in galaxies and quasars \cite{DESI:2024uvr} as well as in Lyman alpha forests \cite{DESI:2024lzq}. \sout{In their analysis, the DESI collaboration noted the preference for the $w_0w_a$-model over $\Lambda$CDM is driven by the luminous red galaxy (LRG) data in the redshift range $0.4 \leq z \leq 0.6$. In some cases in this analysis, we replace select DESI measurements with comparable redshift BOSS DR12 measurements to assess the robustness of the DESI preference for non-cosmological constant dark energy models} \edit{In their analysis, the DESI collaboration noted a preference for a $2\sigma$ deviation from $\Lambda$CDM predictions in the luminous red galaxy (LRG) data in the redshift range $0.4 \leq z \leq 0.6$. Nevertheless DESI found the preference for $w_0w_a$CDM over $\Lambda$CDM persisted, though at weaker significance, when DESI data below $z = 0.6$ were replaced with SDSS data. Similar to the DESI analysis, in some cases in this analysis, we replace select DESI measurements with comparable redshift BOSS DR12 measurements to assess the robustness of the DESI preference for non-cosmological constant dark energy models} \cite{Alam/etal:2017}. In particular, we replace the $D_M/r_d$ and $D_H/r_d$ measurements from DESI in redshift bins $0.4 \leq z \leq 0.6$ ($z_{\rm eff} = 0.51$) and $0.6 \leq z \leq 0.8$ ($z_{\rm eff} = 0.706$) with BOSS DR12 measurements of $D_M/r_d$ and $Hr_d$ at effective redshifts $z_{\rm eff} = 0.51$ and $z_{\rm eff} = 0.61$. This case is referred to as DESI$^{\star}$ + BOSS$_{\rm DR12}^{\star}$.

For supernova, we use four different supernova catalogs to determine how each of these catalogs affects the constraints on cosmological models. In particular, we use supernova from the Pantheon compilation \cite{Scolnic/etal:2018}, the successor analysis Pantheon+ \cite{Brout/etal:2022}, the Union3 analysis \cite{Rubin/etal:2023}, and the Dark Energy Survey Y5 analysis \cite{DES/SNY5}. Importantly, we only include one of these supernova compilations at a time, and we note that these compilations have supernovas in common, which suggests that these results will be correlated with each other. 

% --------------------------------------------------------------------
% --------------------------------------------------------------------
% --------------------------------------------------------------------
\section{Results $\&$ Discussion}
\label{sec:results}
% --------------------------------------------------------------------
% --------------------------------------------------------------------
% --------------------------------------------------------------------

% --------------------------------------------------------------------
% --------------------------------------------------------------------
% --------------------------------------------------------------------
\begin{figure}[!tbp]
\centering
\raisebox{-0.5\height}{\includegraphics[width=0.485\textwidth,angle=0]{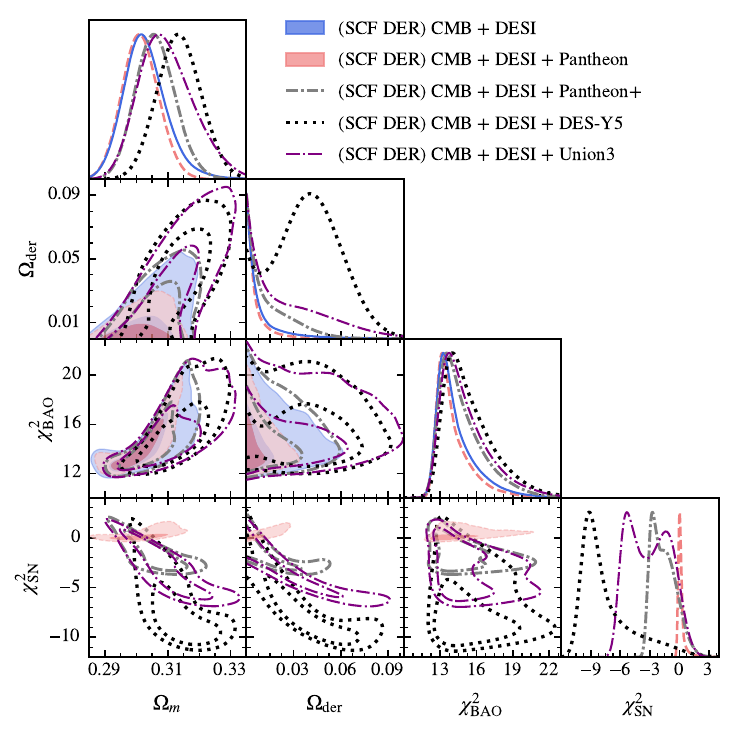}}
\vspace{-0.325cm}
\hfill
\caption{Marginalized posteriors for dark energy radiation (SCF DER), where the dynamical component is dark radiation rather than kinetic energy. The supernova $\chi^2_{\text{SN}}$ is shown with a constant offset. The model's ability to improve the 
fit to supernova data for smaller values of $\Omega_m$ is worse than SCF QUAD, and SCF LIN. The $95\%$ confidence interval includes $\Lambda$CDM ($\Omega_{\text{der}} = 0$) for all of the dataset combinations explored in this work. 
}  
\label{fig:SCF_der}
\end{figure}
% --------------------------------------------------------------------
% --------------------------------------------------------------------
% --------------------------------------------------------------------

% --------------------------------------------------------------------
% --------------------------------------------------------------------
% --------------------------------------------------------------------
\begin{figure*}[!tbp]
\centering
\includegraphics[width=1.0\textwidth,angle=0]{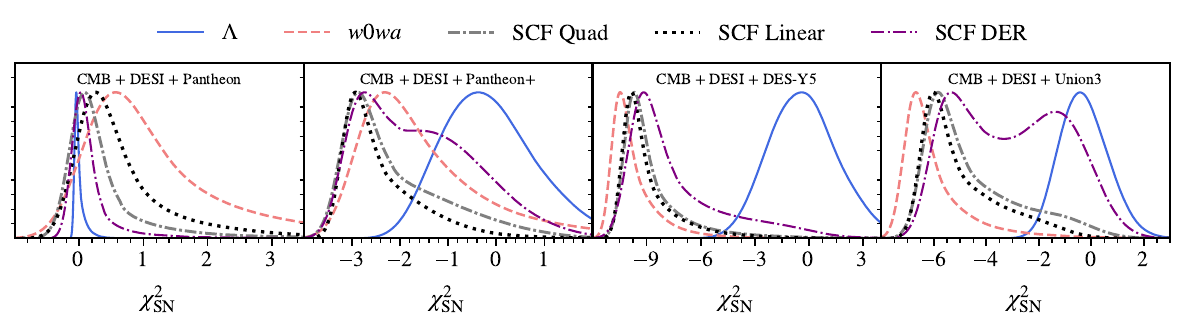}
\includegraphics[width=1.0\textwidth,angle=0]{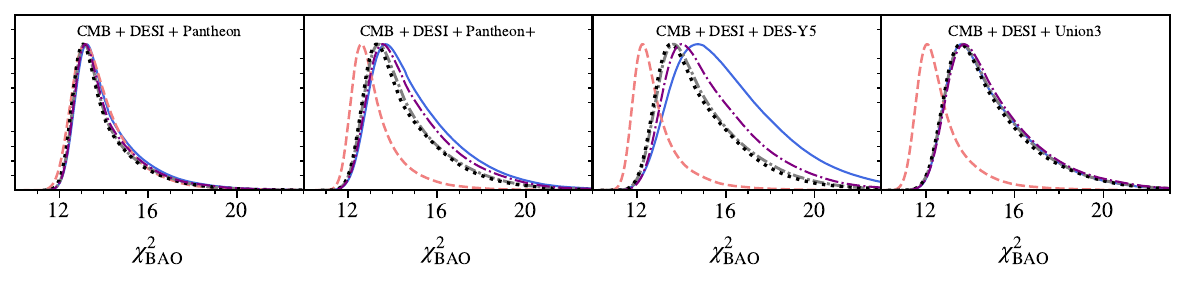}
\hfill
\vspace{-0.5cm}
\caption{ Comparison of the various $\chi^2_{\text{SN}}$ (top panel), and DESI BAO $\chi^2_{\text{BAO}}$ (bottom panel) values, 
between $\Lambda$CDM, $w_0w_a$, and our three scalar field cosmologies. Evolving dark energy cosmologies improve the fit to supernova significantly for Pantheon+, DES-Y5, and Union3. While the peaks of the evolving dark energy models $\chi^2_{\rm{SN}}$ for Pantheon are larger than the $\Lambda$CDM value, the tails of the distributions can achieve a better fit than $\Lambda$CDM. For $\chi^2_{\rm{BAO}}$, the evolving scalar field models perform comparably to $\Lambda$CDM, while the $w_0w_a$ model achieves a slightly lower $\chi^2_{\rm{BAO}}$ value. The supernova $\chi^2_{\text{SN}}$ are shown with the same offset as before.
}
\label{fig:chi2s}
\end{figure*}
% --------------------------------------------------------------------
% --------------------------------------------------------------------
% --------------------------------------------------------------------

We explore the ability of the CMB, BAO, and supernova data outlined in Sec.~\ref{sec:data} to constrain the $\Lambda$CDM, and scalar field dark energy models. In Fig.~\ref{fig:LCDM}, we show posteriors for $\Lambda$CDM when fit to various combinations of CMB, BAO, and supernova data. Fitting $\Lambda$CDM to BAO data in combination with CMB and supernova unveils that every supernova dataset, except for Pantheon, prefers larger values for $\Omega_m$ as evidenced by the negative correlation between $\chi^2_{\rm SN}$ and $\Omega_m$. On the other hand, BAO data favors the opposite regime, indicated by the decreasing $\chi^2_{\text{BAO}}$ for smaller $\Omega_m$. While more pronounced when using the DESI BAO dataset, the trend persists when replacing the two DESI measurements in the redshifts bins for $0.4 \leq z \leq 0.6$ and $0.6 \leq z \leq 0.8$, which were claimed to drive the preference for $w_0w_a$ in \cite{DESI:2024mwx},
with BOSS DR12 measurements, as shown in the right panel of Fig.~\ref{fig:LCDM}. This indicates that the combination of CMB, BAO, and supernova datasets are in mild tension under $\Lambda$CDM, independently of the two redshift bins in question. 

We quantify this tension for $\Lambda$CDM as well as for phenomenological $w_0w_a$ parameterizations in Appendix~\ref{sec:w0wa}. Overall, we find all of the models fit the supernova similarly relative to the number of degrees of freedom for each supernova catalog; however, we find for $\Lambda$CDM, mild to moderate tension (an effective $n_\sigma$\footnote{Note this is not directly interpretable as a number of standard deviations away from the mean. The effective $n_\sigma$ values are calculated using the Tensiometer code, \url{https://github.com/mraveri/tensiometer}. This code uses a two-tailed statistical test, which is outlined in Appendix G of \cite{Raveri/Hu/2019}. For reference, a Probability to Exceed (PTE) of 0.045 corresponds approximately to $2\sigma$. We use this effective $n_\sigma$ throughout the remainder of this work.} $\approx 2-3$) for Pantheon+, DES-Y5, and Union3 between the parameter constraints resulting from MCMCs using these supernova catalogs alone and an MCMC using only CMB and BAO data. This disagreement is reduced when using evolving dark energy models, which motivates exploring the constraints of our scalar field cosmologies.

The results for our scalar field dark energy cosmologies are shown in Fig.~\ref{fig:SCF} (SCF QUAD and SCF LIN) and Fig.~\ref{fig:SCF_der} (SCF DER). Additionally, we provide select parameter constraints for these cosmologies in Table~\ref{tab:parameter_table1}. The $\chi^2_{\text{SN}}$ row in Fig.~\ref{fig:SCF} shows that the hypothesis of thawing scalar fields with a quadratic or linear potential is able to ease the preference for larger $\Omega_m$ from the supernova datasets as seen by the allowed parameter space with $\chi^2_{\rm SN} < 0$ and $\Omega_m < 0.31$. This opening of parameter space helps accommodate the BAO preference for lower $\Omega_m$, which persists even in these scalar field dark energy models, partially because the $\chi^2_{\rm BAO}-\Omega_m$ posterior is less steep (i.e., smaller shifts in $\chi^2_{\rm BAO}$ as $\Omega_m$ increases) when $\Omega_m < 0.31$.

%\jk{Should we cut this next sentence? I think I added it earlier and I'm not sure how much it is adding at this point. At the very least we need to get rid of the reference to the Pantheon+ case because it is not true.} Interestingly, Table~\ref{tab:parameter_table1} shows that for SCF QUAD and SCF LIN CMB + DESI, the mean $\Omega_m$ values are higher than the mean values from CMB + DESI + Pantheon and CMB + DESI + Pantheon+. 

Moreover, the $\Omega_{\rm scf,k}$ rows, for both SCF QUAD and SCF LIN cases, \edit{show weak statistical evidence} for nonzero kinetic scalar field energy. 
%This preference is strongest 
\edit{The mean values are largest} for the cases where the DES-Y5 and Union3 compilations are used, which for both a quadratic and linear scalar field are around  $4\%$ kinetic energy. For Pantheon and Pantheon+, the preference for nonzero kinetic energy is weaker, around $2\%$.
In all cases explored, except 
the DES-Y5 dataset, the 1D $95\%$  confidence intervals for the scalar field kinetic energy \edit{includes the lower boundary allowed by our prior, which indicates the $\Lambda$CDM limit of the model. }

In Fig.~\ref{fig:SCF_der}, we show that the SCF DER model also allows for a lowering of the $\chi^2_{\rm SN}$ for $\Omega_m < 0.31$.
Again, DES-Y5 and Union3 show more pronounced hints of dynamical dark energy than Pantheon or Pantheon+. While the $2\sigma$ upper limits for the DES-Y5 and Union3 combinations are compatible with up 
$7-8\%$ of the energy of the universe being comprised of dark energy radiation, the lower boundary of all contours are compatible with the $\Lambda$CDM limit (i.e., $\Omega_{\text{der}} = 0$).

% --------------------------------------------------------------------
% --------------------------------------------------------------------
% --------------------------------------------------------------------
\begin{figure}[!tbp]
\centering
\hspace{-0.8cm}
\raisebox{-0.5\height}{\includegraphics[scale = 0.99,angle=0]{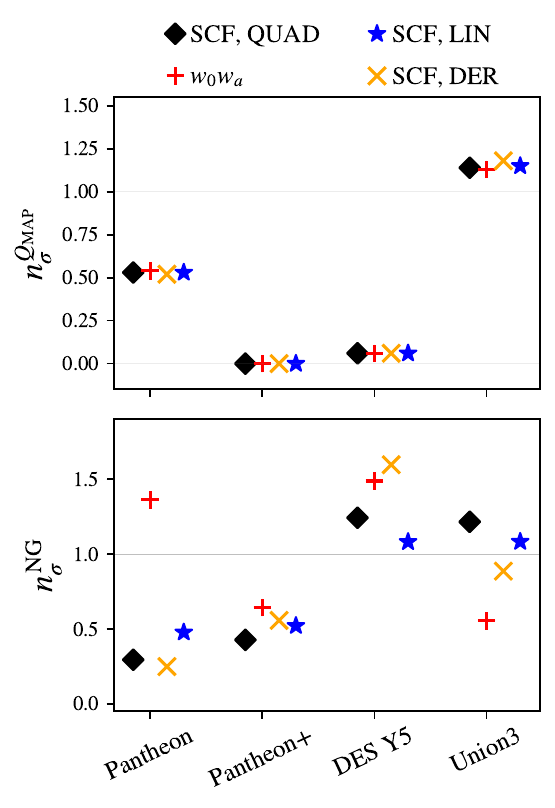}}
\vspace{-0.325cm}
\hfill
\caption{ (Top) Plot of the $Q_{\rm{MAP}}$ goodness of fit statistic for each supernova compilation. In each case, we find the best-fit point from an MCMC chain that includes CMB, BAO, and supernova, calculate the effect number of degrees of freedom for the experiment, and compare the goodness of fit of the $\chi^2_{\rm{SN}}$ at the overall best-fit to this number of degrees of freedom. We then quantify the goodness of fit in terms of an effective number of $\sigma$. In all cases, the models are consistent with measured supernova data. (Bottom) Plot of the cosmological parameter difference tension between posteriors resulting from supernova alone and posteriors resulting from CMB + BAO. Because the posteriors are, in general, non-Gaussian, we use a non-Gaussian tension metric (Kernel Density Estimates) and quantify this in terms of the same effective number of $\sigma$ as in the top plot. Each of the evolving dark energy models results in less than $2\sigma$ tension between parameter determinations. 
}  
\label{fig:tension}
\end{figure}
% --------------------------------------------------------------------
% --------------------------------------------------------------------
% --------------------------------------------------------------------

Fig.~\ref{fig:chi2s} compares the $\Lambda$CDM, $w_0w_a$, SCF QUAD, SCF LIN, and SCF RAD $\chi^2_{\rm SN}$ and $\chi^2_{\rm BAO}$ values for each of the supernova data combinations. In all cases, the extended cosmological models fit the supernova data better than $\Lambda$CDM, though note that for the Pantheon case, this reduction in $\chi^2_{\rm SN}$ is only in the tails of the distributions as the peaks for the extended cosmologies are at greater $\chi^2_{\rm SN}$ than the $\Lambda$CDM case. For the Pantheon+, DES-Y5, and Union3 cases, the evolving dark energy models fit the supernova data better than $\Lambda$CDM with $\Delta \chi^2_{\rm SN}$ values roughly equal to -3, -9, and -6 respectively. 

For the Pantheon, Pantheon+ cases, and DES-Y5 the scalar field models fit the corresponding supernova data better than the phenomenological $w_0w_a$ model, though this is reversed for Union3. In all cases, the scalar field dark energy models can fit the supernova data almost as well as the $w_0w_a$ model. The peaks of the $\chi^2_{\rm SN}$ distributions for the SCF QUAD and SCF LIN are lower than SCF DER for the Pantheon+, DES-Y5, and Union3 cases. 

For the fits to the BAO data, all models have at least a comparable fit to the BAO data as the $\Lambda$CDM case. However, there is a slight worsening in the case when Union3 supernova data are included for the scalar field models. In all cases, the $w_0w_a$ model fits the BAO data better than all other models explored. This highlights that while most of the improvement to the combined BAO and supernova fits from dynamical dark energy models comes from an improvement in the fit to the supernova data, there is still some room for improvement in the fit to the BAO as well. 

% --------------------------------------------------------------------
% --------------------------------------------------------------------
% --------------------------------------------------------------------
\begin{table}
\centering
\resizebox{\columnwidth}{!}{%
    \small
    \begin{tabular}{lccccc}
    \toprule
    \midrule
    \multirow{2}{*}{model/dataset} & \multirow{2}{*}{$\Omega_m$} & \multirow{2}{*}{$\Omega_{\text{scf,k}}$} & \multirow{2}{*}{$\Omega_{\text{der}}$} \\ \\
    \midrule
    \bf{$\Lambda$CDM} &&&&\\
    DESI+CMB & $0.300 \pm 0.011 $ & ---  & --- & \\
    DESI+CMB+Panth. &  $0.300^{+0.011}_{-0.010}$  & --- &  ---& \\
    DESI+CMB+Panth.+ & $0.303 \pm 0.011$ & --- & ---&\\    
    DESI+CMB+Union3 & $0.303 \pm 0.011$ & ---  & ---&\\
    DESI+CMB+DESY5 & $0.306 \pm 0.011$ & --- & ---&\\
    \midrule
    \bf{SCF QUAD} &&&&\\
    DESI+CMB & $0.307^{+0.021}_{-0.017}$ & $0.017 \ [0,0.062]$  & ---& \\
    DESI+CMB+Panth. &  $0.302 \pm 0.012$  & $0.008 \ [0, 0.024]$ &  ---& \\
    DESI+CMB+Panth.+ & $0.308^{+0.013}_{-0.012}$ & $0.018 \ [0, 0.040]$ & ---&\\    
    DESI+CMB+Union3 & $0.315 \pm 0.018$  & $0.035 \ [0, 0.071]$ & ---&\\
    DESI+CMB+DESY5 & $0.316 \pm 0.013$ & $0.037^{+0.027}_{-0.029}$ & ---&\\
    \midrule
    \bf{SCF LIN} &&&&\\
    DESI+CMB  & $0.310^{+0.023}_{-0.019}$ & $0.028 \ [0, 0.076]$ & ---& \\
    DESI+CMB+Panth. & $0.304 \pm 0.012$  & $0.013 \ [0, 0.030]$ &  ---& \\
    DESI+CMB+Panth.+ & $0.309 \pm 0.013$  & $0.022 \ [0,0.044]$ & ---&\\    
    DESI+CMB+Union3 & $0.316 \pm 0.018$ & $0.042 \ [0, 0.081]$ & ---&\\
    DESI+CMB+DESY5 & $0.316 \pm 0.013  $ & $0.040 \pm 0.29$ & ---&\\    \midrule    
    \bf{SCF DER} &&&&\\
    DESI+CMB & $0.303 \pm 0.014$  &---& $0.001 \ [0,0.039]$ \\
    DESI+CMB+Panth. & $0.301^{+0.012}_{-0.011}$  &---& $0.006 \ [0,0.022]$ \\
    DESI+CMB+Panth.+ & $0.306^{+0.013}_{-0.012}  $  &---& $0.015 \ [0,0.043]$ \\    
    DESI+CMB+Union3 & $0.310^{+0.018}_{-0.016}$  &---& $0.027 \ [0,0.076]$ \\
    DESI+CMB+DESY5 & $0.314^{+0.014}_{-0.013}$ &---& $0.039 \ [0,0.073]$ \\
    \midrule
    \bottomrule
    \end{tabular}
    }
\caption{
    The mean and 1D 95$\%$ confidence intervals of the cosmological parameters, $\Omega_m$, and the dynamical scalar field component $\Omega_{\text{scf,k}}$ or  $\Omega_{\text{der}}$ for the considered scalar field dark energy cosmologies. 
    }  \vspace{0.1em}
    \label{tab:parameter_table1}
\end{table}
% --------------------------------------------------------------------
% --------------------------------------------------------------------
% --------------------------------------------------------------------

Interestingly, this improvement in the fit to the BAO data is not found in the scalar field cases. In Appendix~\ref{sec:w0wa}, we study the phenomenology of $w_0w_a$ parameterizations where we restrict either the $w_0$ or $w_a$ parameter space to correspond to thawing, tracking, or phantom dark energy cases, respectively. Similar to the scalar field models, these phenomenological models improve the fit to supernova data relative to $\Lambda$CDM, but do not improve the fit to BAO data.  
 
% --------------------------------------------------------------------
% --------------------------------------------------------------------
% --------------------------------------------------------------------
\begin{figure*}[!tbp]
\centering
\includegraphics[width=\textwidth,angle=0]{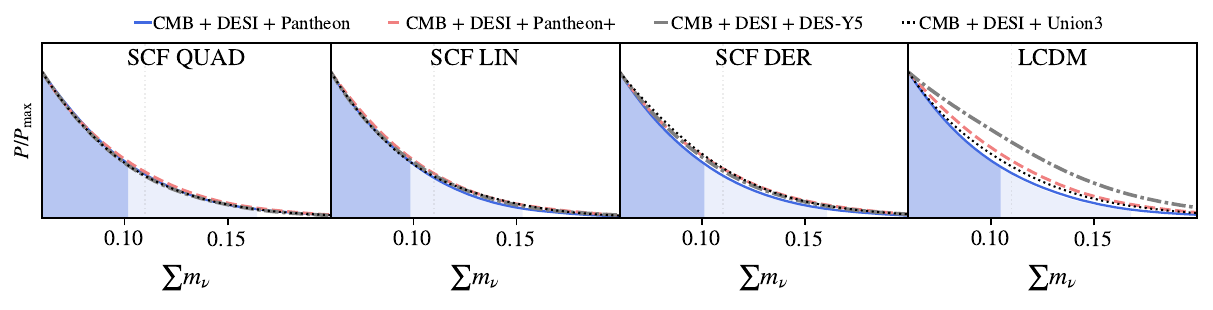}
\vspace{-0.5cm}
\hfill
\caption{ To this point, we have fixed $\Sigma m_{\nu}$. In this plot, we test the impact of allowing $\Sigma m_{\nu}$ to vary for each of the scalar field dark energy models explored in this work and for each supernova compilation. The dark (light) bands correspond to the 68$\%$ (95$\%$) confidence intervals. For all cases, the constraints on $\Sigma m_{\nu}$ are comparable to those obtained by the DESI collaboration assuming $\Lambda$CDM. The sum of neutrino masses is in $\text{eV}$. 
}  
\label{fig:neutrino}
\end{figure*}
% --------------------------------------------------------------------
% --------------------------------------------------------------------
% --------------------------------------------------------------------

In the top panel of Fig.~\ref{fig:tension}, we quantify the goodness of fit of each of the SCF QUAD, SCF LIN, SCF DER, and $w_0w_a$ models to the supernova data using the $Q_{\rm{MAP}}$ test in \cite{Raveri/Hu/2019}. In this test, we evaluate the goodness of fit of the models to the supernova data given the CMB and BAO constraints as a prior. In particular, we find the best-fit point from MCMC runs of CMB + BAO + SN, which are referred to as the maximum a posteriori (MAP) point. We take the corresponding supernova likelihood $\chi^2$ at the MAP point and compare it to the effective number of degrees of freedom from the supernova compilation, taking into account the number of parameters that the supernova data actually constrain given the CMB and BAO data already provide constraints on parameters. 

For DES-Y5, we use 1735 as the number of effective data points instead of the total number of supernova data points 1829 as recommended by \cite{DES/SNY5} to account for a slight overestimation of uncertainties in their data covariance matrix. For Pantheon+, we use 1590 for the number of data points, which corresponds to the number of data points in the Pantheon+ sample when excluding the SH0ES supernova. 
%We note that the Pantheon+ data covariance matrix is likely overestimating errors because of relatively low $\chi^2_{\rm{SN}}$ values ($\approx 1400$) relative to the total number of supernova light curves, 1701, but we use the 1701 number for the number of data points as we do not know how to account for this potential overestimation. 

In all cases, we find the models are consistent with the supernova data with $n_{\sigma} < 2$. In general, the variations in the $n_{\sigma}^{Q_{\rm{MAP}}}$ values between models are smaller than the variations between supernova data sets. This is generally because the changes in $\chi^2$ are small relative to the number of data points, which for Pantheon, Pantheon+, and DES-Y5 are each over 1000. In Appendix~\ref{sec:w0wa}, we also show the results for $\Lambda$CDM, which are comparable to these results except for Union3 where $n_{\sigma}^{Q_{\rm{MAP}}} = 1.7$. This shift is larger for Union3 because their data have only 22 bins. 

In the bottom panel of Fig.~\ref{fig:tension}, we quantify the tension between parameter constraints from supernova alone compared to parameter constraints from CMB + DESI data alone. For the tension tests, we always use parameters $\Omega_m$ and $H_0$. For $w_0w_a$, we also include $w_0$ and $w_0+w_a$, while for the scalar field models, we use the single parameter extension in each case. Because we are exploring extended cosmologies and using only these data subsets, the resulting posteriors are non-Gaussian. We tried using simple Gaussian methods to quantify the level of agreement between the parameter constraints from supernova and the parameter constraints from CMB + DESI, and we found that these resulted in evidence for weak to mild tension (up $\approx 2\sigma$ in the most extreme case). 

To account for the non-Gaussianity in the parameter posteriors, we use the Tensiometer code described above. In particular, we use the Kernel Density Estimates (KDE) of the parameter difference posterior as described in \cite{Raveri/etal:2021}. For each of these evolving dark energy model cases, the effective $n_{\sigma} < 2$ indicates increased compatibility between the supernova and CMB + DESI determinations of the parameter constraints. 
This number can be compared directly to the level of tension
when assuming $\Lambda$CDM where in the most extreme case $n_{\sigma}^{NG} = 3.4 \, (2.8)$ for DESI + DES-Y5 (BAO = DESI$^{*}$+BOSS$^{*}_{\rm DR12}$ ), but also takes on values of $1.7 \, (1.4)$ and $2.1 \, (1.9) $ for Pantheon+ and Union3,  respectively. 
In all three of these cases, the evolving dark energy models reduce the level of tension between parameter determinations using only supernova vs parameter determinations using CMB + DESI. For the Pantheon case, the evolving dark energy models have comparable if not slightly higher $n_{\sigma}^{NG}$ values than $\Lambda$CDM. Overall, the scalar field models have approximately the same $n_{\sigma}^{NG}$ values, and in some cases lower, as the $w_0w_a$ model despite not including a phantom crossing. 
%In particular, the quadratic scalar field model provides the best fit of the scalar field models for Pantheon+, DES-Y5, and Union3, outperforming $w_0w_a$ for Pantheon+, and DES-Y5, and in all cases having an effective $n^{\text{NG}}_{\sigma} <0.7$. 
 
In light of the constraining power of DESI BAO measurements on the sum of the neutrino masses, we consider the impact of our scalar field cosmologies on the constraints of the sum of the neutrino masses.
The minimal sum of neutrino masses based on results from neutrino oscillation experiments amounts to $\sum {m_{\nu}} = 0.11\, \text{eV}$, in the inverted mass hierarchy and to $\sum m_{\nu} = 0.059 \, \text{eV}$. 
Combined with the $\Lambda$CDM model of cosmology, the DESI BAO measurements favor a normal neutrino hierarchy over an inverted one. The bounds relax noticeably under the $w_0w_a$-parameterization. We quantify the relaxation of the bounds for the three scalar field cosmologies we consider. We adopt the standard neutrino description with one massive ($m_{\nu} = \sum m_{\nu}$ eV) and two massless neutrinos, and marginalize over the neutrino mass sum with priors $\sum m_{\nu} \in [0.6,0.3]$\footnote{The DESI BAO results use a description of three degenerate neutrinos which differs from ours.}. We present the results in Fig.~\ref{fig:neutrino}.
As has been noted before \cite{Vagnozzi:2018jhn}, the bounds from our scalar field cosmologies are comparable to the $\Lambda$CDM ones, as the neutrino bounds weaken significantly only if one allows for a phantom crossing in the dark energy equation of state.

% --------------------------------------------------------------------
% --------------------------------------------------------------------
% --------------------------------------------------------------------
\section{Conclusions}
\label{sec:concl}
% --------------------------------------------------------------------
% --------------------------------------------------------------------
% --------------------------------------------------------------------

In this work we have evaluated three simple scalar field models as candidates for evolving dark energy in the light of the BAO DESI Y1 release. We find that any preference for evolving dark energy is not only driven by the low redshift BAO DESI data, but persists when replacing them with comparable measurements from BOSS DR 12, in combination with CMB and supernova data sets. 
We pinpoint the source of the preference to be due to BAO preferring smaller values for $\Omega_m$ under $\Lambda$CDM, while every supernova data set, except for Pantheon, prefers larger values of $\Omega_m$. Evolving dark energy reduces the mild tension under $\Lambda$CDM ($n_{\sigma} <2 -3.4$) by achieving a better fit to supernova data with smaller $\Omega_m$, which are preferred by BAO. 

Furthermore, we find that the simplest canonical scalar field models fit the data well. Notably, in particular the quadratic and linear scalar field models are discrepant with the $\Lambda$CDM model at the $95\%$ confidence level in combination with the DES-Y5 supernova dataset.
%for every supernova data set combination,
%except for Pantheon, 
%highlighting the preference for dynamics.
The preferred cosmologies feature kinetic scalar field energies comprising between $2-4\%$ of the energy density of the universe today. For the dark energy radiation hypothesis, the best fit ranges from $1-4\%$ of dark energy radiation.
%Highlighting the scalar field with a quadratic potential, the model succeeds in reducing the tension between BAO + CMB and supernova data sets below $n_{\sigma}< 1$, outperforming $w_0w_a$.
We conclude that simple canonical scalar field models are able to successfully mitigate the mild tension
under $\Lambda$CDM, demonstrating that evolving dark energy with a phantom crossing is not required to explain the DESI BAO measurements. In the absence of such a phantom crossing, the bounds on the neutrino mass sum remain comparable to $\Lambda$CDM, even for evolving scalar field dark energy.  

% --------------------------------------------------------------------
% --------------------------------------------------------------------
% --------------------------------------------------------------------
\section*{Acknowledgments}
% --------------------------------------------------------------------
% --------------------------------------------------------------------
% --------------------------------------------------------------------
We sincerely thank Elisabeth Krause and Tim Eifler for providing computational resources that allowed us to complete this work on an ambitious timeline. We thank Ryan Camilleri for porting the DES-Y5 SN likelihood to \textsc{cobaya}, and David Rubin for providing the Union3 SN likelihood files. We also thank Antony Lewis for his incredible work porting the recent DESI and SN results to \textsc{cobaya}. We also thank Dillon Brout for useful discussions.
K.B. thanks the U.S. Department of Energy, Office of Science, Office of High Energy Physics, under Award Number DE-SC0011632, and the Walter Burke Institute for Theoretical Physics.
The simulations in this paper use High-Performance Computing (HPC) resources supported by the University of Arizona TRIF, UITS, and RDI and maintained by the UA Research Technologies department. 
The authors would also like to thank the Stony Brook Research Computing and Cyberinfrastructure, and the Institute for Advanced Computational Science at Stony Brook University for access to the high-performance SeaWulf computing system, which was made possible by a $\$1.4$M National Science Foundation grant ($\# 1531492$).

% --------------------------------------------------------------------
% --------------------------------------------------------------------
% --------------------------------------------------------------------
\begin{figure*}[!tbp]
\centering
\includegraphics[width=1.0\textwidth,angle=0]{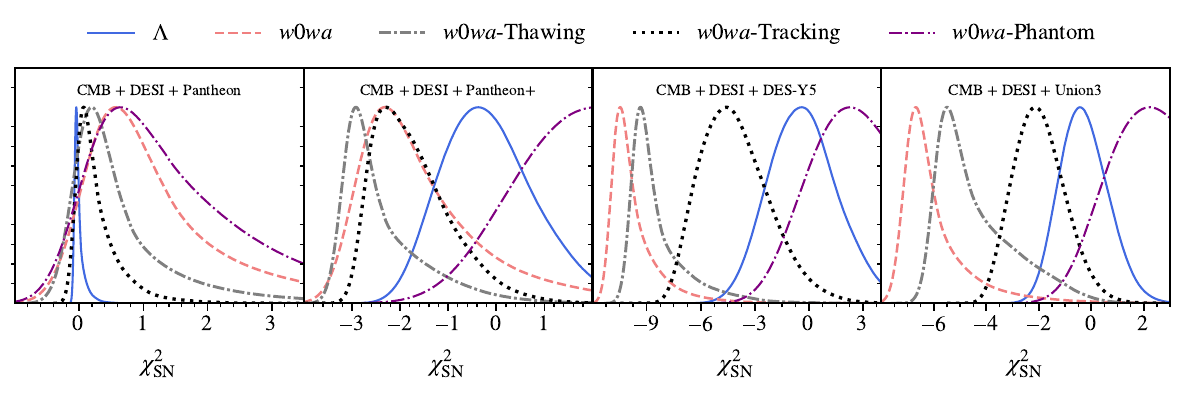}
\vspace{0.4cm}
\includegraphics[width=1.0\textwidth,angle=0]{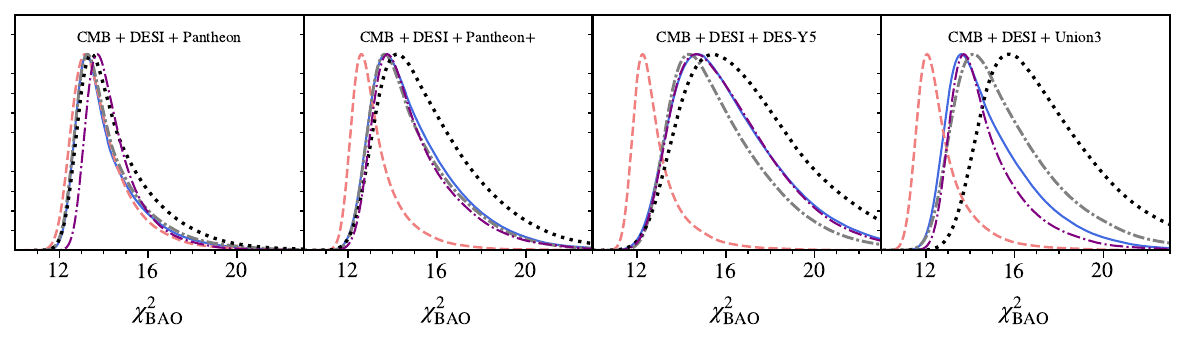}
\includegraphics[width=1.0\textwidth,angle=0]{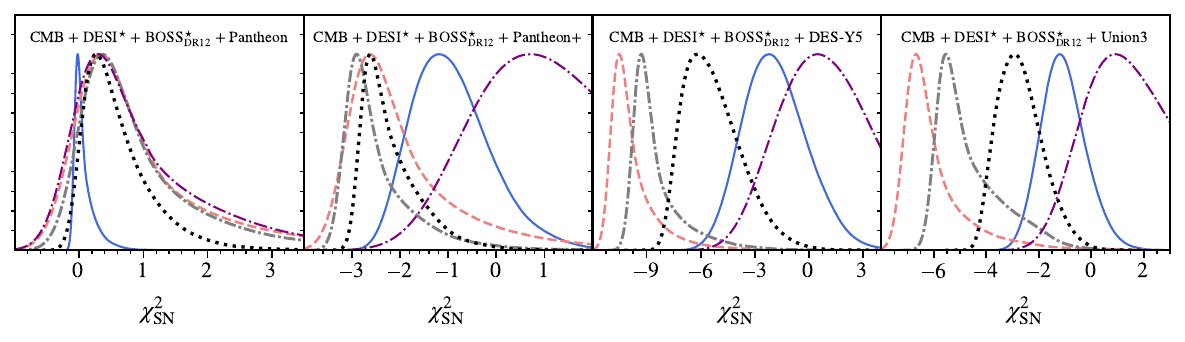}
\includegraphics[width=1.0\textwidth,angle=0]{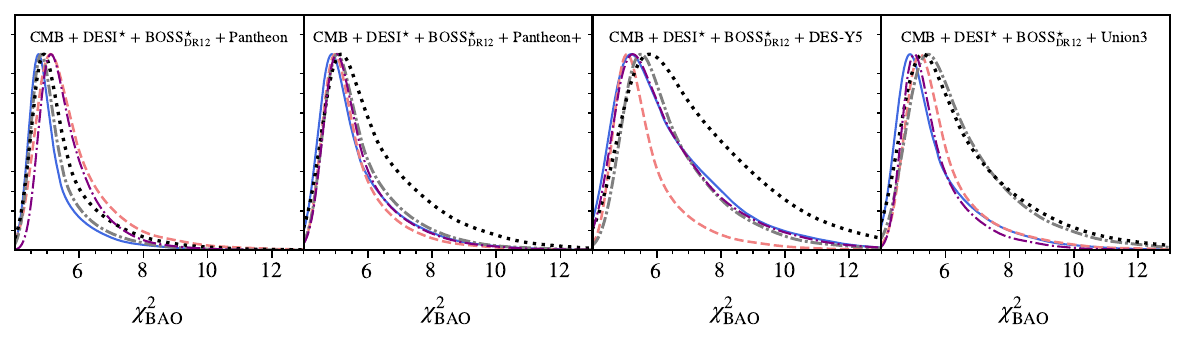}
\hfill
\vspace{-0.6cm}
\caption{Comparison of supernova $\chi^2_{\text{SN}}$ , and BAO $\chi^2_{\text{BAO}}$, 
between $\Lambda$CDM, $w_0w_a$, $w_0w_a$-thawing, $w_0w_a$-tracking, and $w_0w_a$-phantom for two different combinations of BAO data. Similar to scalar field cases, the DES-Y5 and Union3 data prefer evolving dark energy, while DESI BAO only prefers $w_0w_a$ over $\Lambda$CDM. $w_0w_a$-phantom generally fits the data worse than $\Lambda$CDM.  }  
\label{fig:w0wa1}
\end{figure*}
% --------------------------------------------------------------------
% --------------------------------------------------------------------
% --------------------------------------------------------------------

% --------------------------------------------------------------------
% --------------------------------------------------------------------
% --------------------------------------------------------------------
\begin{figure*}[!tbp]
\centering
\includegraphics[width=0.485\textwidth,angle=0]{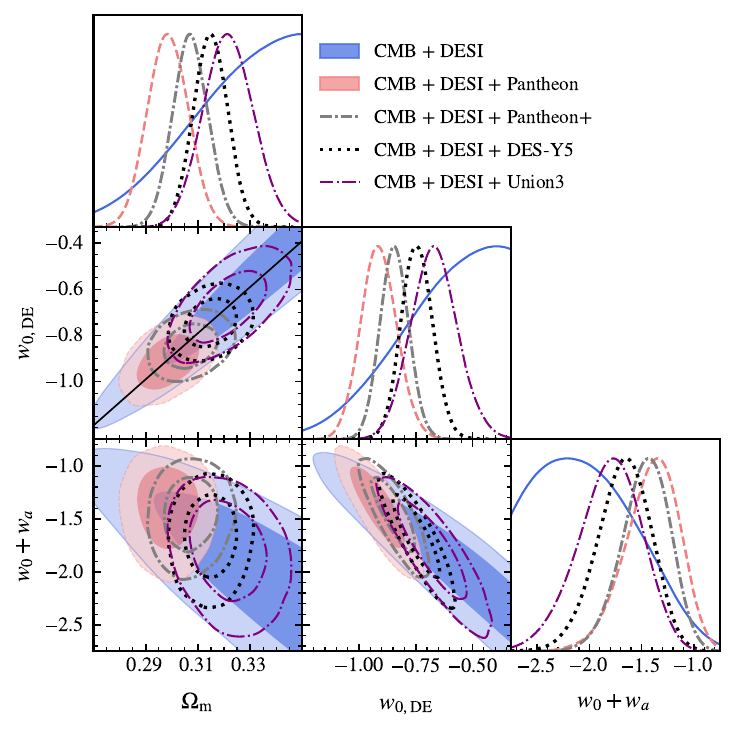}
\includegraphics[width=0.485\textwidth,angle=0]{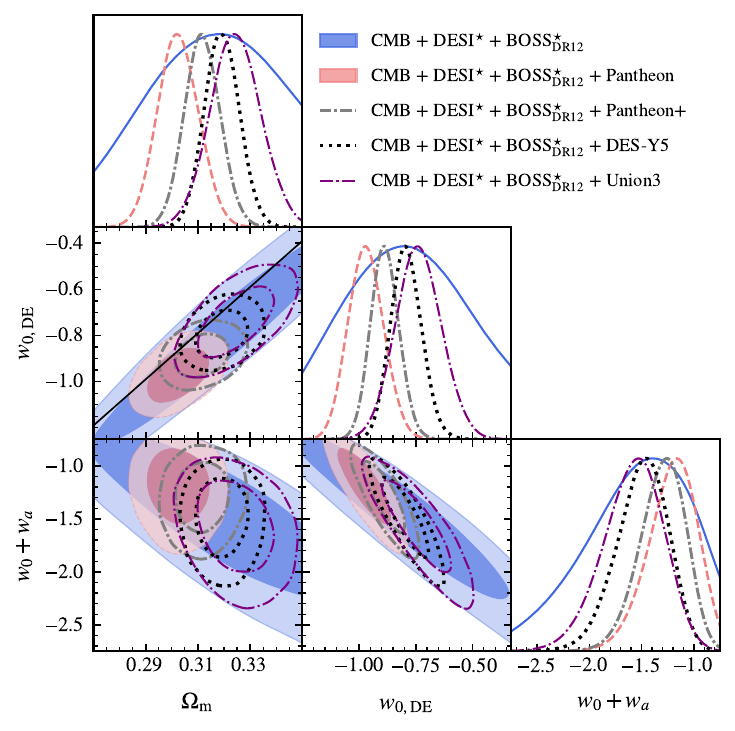}
%x
\caption{Marginalized posteriors for $w_0w_a$ for different supernova datasets. On the left, we combine with DESI BAO data. On the right, we replaced two of the DESI redshifts claimed to drive the preference at $z_{\rm eff} =0.51$, and $z_{\rm eff} =0.706$  for evolving dark energy with BAO measurements from BOSS DR12 measurements at $z_{\rm eff} =0.51$ and $z_{\rm eff} =0.61$, as described in detail in Sec.~\ref{sec:data}. The black line along the $w_0\Omega_m$-degeneracy direction on the left plot is given by $w_0 = -1 + 10 \times(\Omega_m-0.289)$. We use the same degeneracy line in the right plot where we replace the two DESI redshifts. In this cases, The PC amplitude shifts, which is further illustrated in Fig.~\ref{fig:alpha}. }
\label{fig:w0wa2}
\end{figure*}
% --------------------------------------------------------------------
% --------------------------------------------------------------------
% --------------------------------------------------------------------

% --------------------------------------------------------------------
% --------------------------------------------------------------------
% --------------------------------------------------------------------
\begin{figure*}[!tbp]
\centering
\includegraphics[width=0.485\textwidth,angle=0]{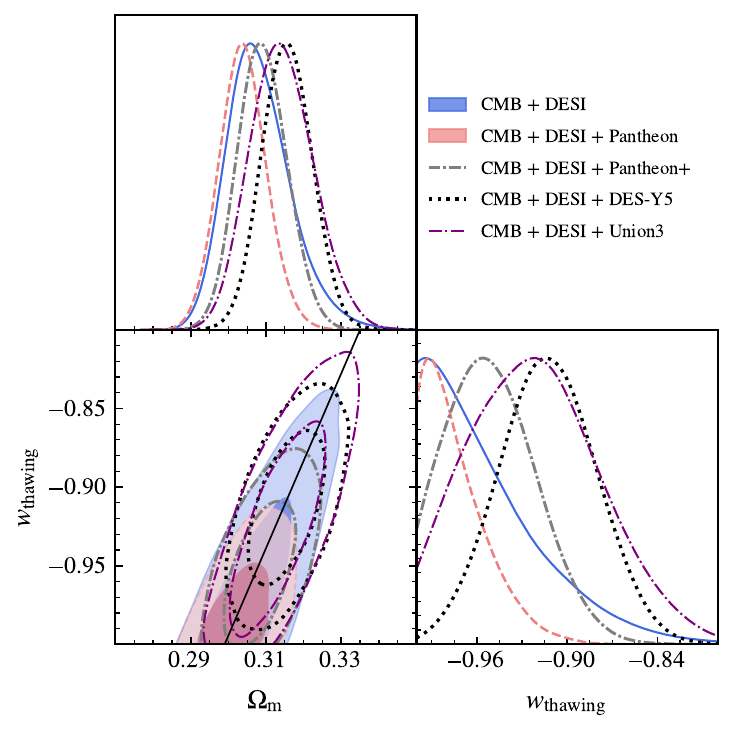}
\includegraphics[width=0.485\textwidth,angle=0]{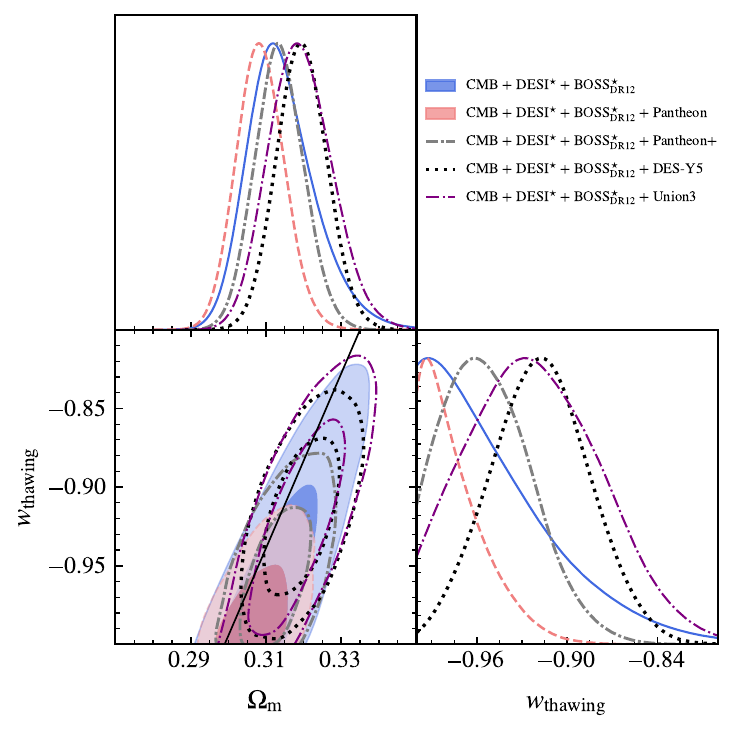}
\hfill
\caption{Marginalized posteriors for $w_0w_a$-thawing for DESI BAO data on the left, and DESI in combination with BOSS DR12, where we have replaced the DESI redshift bins $z_{\rm eff} =0.51$, and $z_{\rm eff} = 0.61$ with BOSS DR12 measurements for all supernova datasets. The black line along the $w_\text{thawing}\Omega_m$-degeneracy direction is given by $w_{\text{thawing}}=-1+(\Omega_m-0.298)/0.18$ for the left plot. For the right plot, the posteriors shift orthogonally to this degeneracy direction.} 
\label{fig:w0wathawing}
\end{figure*}
% --------------------------------------------------------------------
% --------------------------------------------------------------------
% --------------------------------------------------------------------

\section{}
\appendix
\section{$w_0 w_a$ parameterizations}
\label{sec:w0wa}
The $w_0w_a$-model is defined by parameterizing the equation of state of dark energy as 
\begin{equation}
    w(z)  = w_0 + \left(1-\frac{1}{1+z}\right) w_a \,.
\end{equation}
 We consider the standard $w_0w_a$-parameterization with priors $w_0 \in [-3, -0.01]$ and $w_0+w_a \in [-5, 0.01]$. We explore three additional $w_0w_a$-parameterizations that approximate the behavior of tracking, thawing, and phantom scalar fields. For these three models we choose the priors $w_0 = w_{\text{thawing}} \in [-1,-0.01]$, and $w_0 + w_a \in [-1,-0.98]$, $w_0 = w_{\text{tracking}} \in [-1,-0.98]$, and $w_0 + w_a \in [-1,0.01]$, and $w_0 = w_{\text{phantom}} \in [-3,-1]$, and $w_0 + w_a \in [-3,-1]$. The best fits obtained by the DESI BAO data release for a $w_0w_a$ cosmology imply a preference for phantom crossing from $w(z) > -1$ to $w(z)< -1$, which is not a subset of the priors imposed for $w_0w_a$-tracking, $w_0w_a$-thawing, or $w_0w_a$-phantom. For perturbations, we follow the default treatment in CLASS and CAMB, detailed in \cite{Fang:2008sn, Ballesteros:2010ks}. However, we find that our analysis is dominated by the impact of the considered model of the dark energy's equation of state.

% --------------------------------------------------------------------
% --------------------------------------------------------------------
% --------------------------------------------------------------------
\begin{figure}[!tbp]
\centering
\includegraphics[width=0.414\textwidth,angle=0]{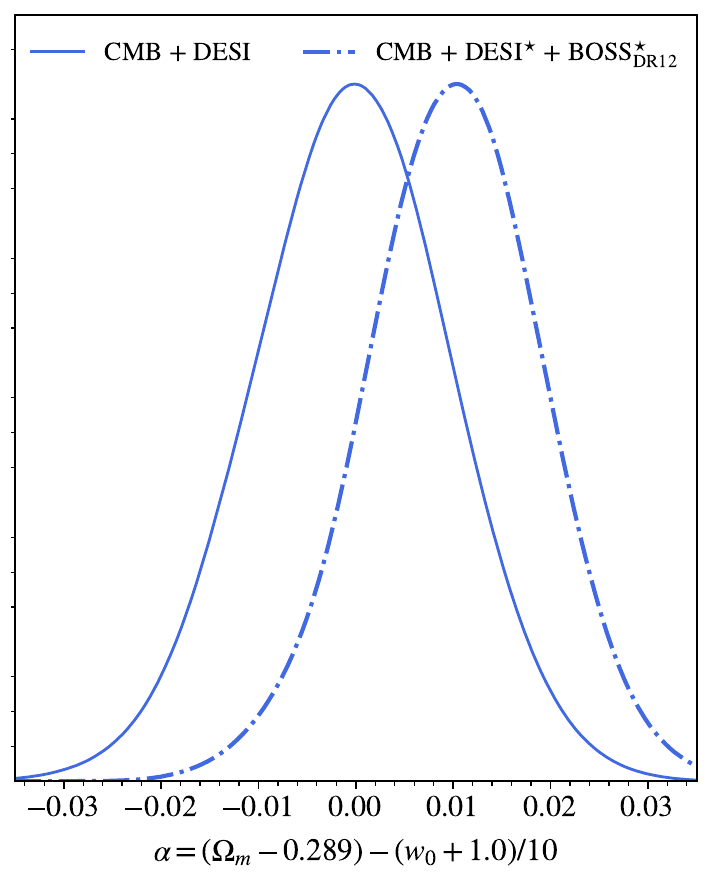}
\includegraphics[width=0.42\textwidth,angle=0]{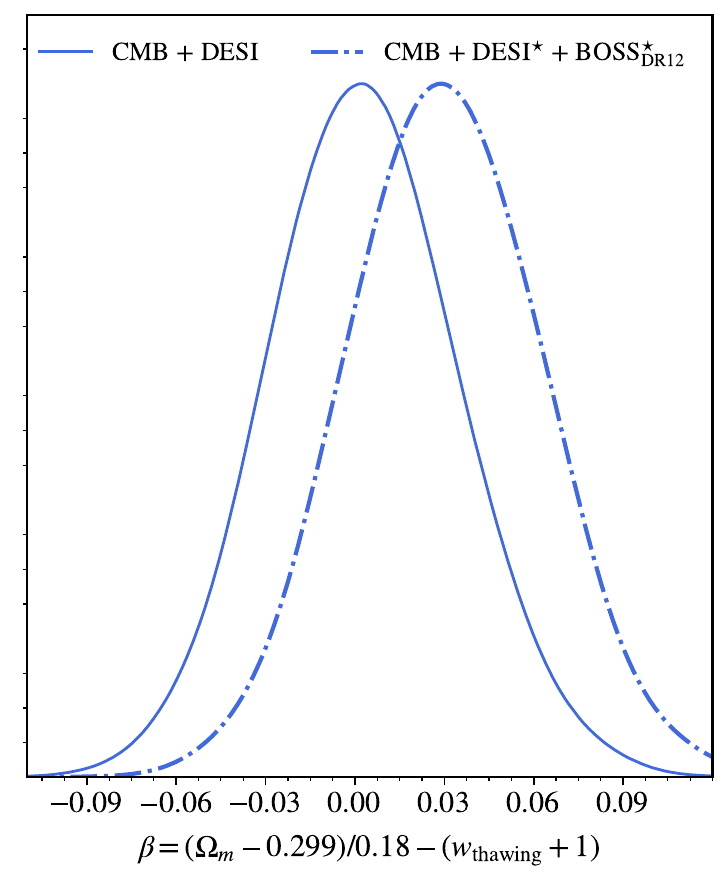}
\vspace{-0.35cm}
\caption{Comparison of the PC amplitude for two BAO data combinations for $w_0 w_a$ (top) and $w_0 w_a$-thawing (bottom). Replacing the two DESI data points at $z_{\rm eff} =0.51$, and $z_{\rm eff} = 0.61$ shifts the value of $\Omega_m$ compatible with $\Lambda$CDM, specified in \eqref{eq:A2}, and \eqref{eq:A3} by the PC amplitude $\alpha$ for $w_0w_a$, and $\beta$ for $w_0w_a$-thawing. 
} 
\label{fig:alpha}
\end{figure}
% --------------------------------------------------------------------
% --------------------------------------------------------------------
% --------------------------------------------------------------------

% --------------------------------------------------------------------
% --------------------------------------------------------------------
% --------------------------------------------------------------------
\begin{figure}[!tbp]
\centering
\hspace{-1.5cm}
\raisebox{-0.5\height}{\includegraphics[scale = 1,angle=0]{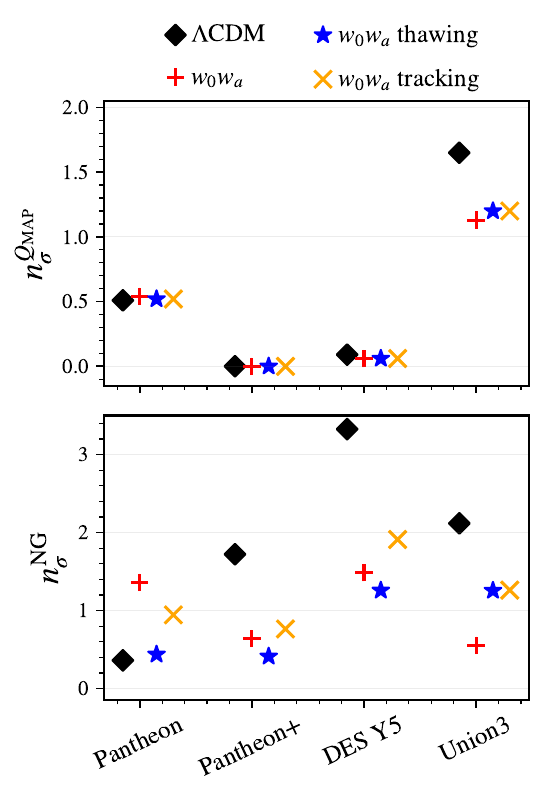}}
\vspace{-0.325cm}
\hfill
\caption{ Same as Figure~\ref{fig:tension}, in the top plot we quantify the $Q_{\rm{MAP}}$ goodness of fit statistic for each supernova compilation given the CMB and BAO data, and in the bottom plot we quantify the non-Gaussian parameter difference tension between the CMB + BAO and the supernova data. As in Figure~\ref{fig:tension}, all of the models result in good fits to the measured data and are consistent with statistical fluctuations (i.e. $n_{\sigma}^{Q_{\rm{MAP}}} < 2$). However, for the $\Lambda$CDM case, there are mild to moderate tensions (up to $n_{\sigma}^{Q_{\rm{MAP}}} \approx 2-3$) between the parameter determinations for supernova alone versus CMB + BAO alone. This is reduced when the evolving dark energy models are used instead of $\Lambda$CDM. This pattern behavior is reversed for Pantheon data where the $\Lambda$CDM case corresponds to the case where the parameters are most consistent. }
\label{fig:tension2}
\end{figure}
% --------------------------------------------------------------------
% --------------------------------------------------------------------
% --------------------------------------------------------------------

In Fig.~\ref{fig:w0wa1}, we show a comparison of $\chi^2$ values for supernova and BAO data, comparing $\Lambda$CDM with $w_0w_a$, $w_0w_a$-thawing, $w_0w_a$-tracking, and $w_0w_a$-phantom. We show the results for both of our BAO dataset combinations.
Fig.~\ref{fig:w0wa1} illustrates that thawing behavior is able to effectively minimize the supernova $\chi^2_{\text{SN}}$ compared to tracking and phantom, as well as $\Lambda$CDM. In fact, for the Pantheon+ dataset, the thawing parameterization even performs better than $w_0w_a$. However, $w_0w_a$ improves the BAO $\chi^2$ over $\Lambda$CDM, in particular for the DESI dataset. That improvement is less pronounced for the DESI$^*$ + BOSS$_{\rm{DR12}}^{*}$, where we have replaced the outlier DESI data points with BOSS measurements, showing that DESI penalizes $w_0w_a$-thawing less.

In Fig.~\ref{fig:w0wa2}  and Fig.~\ref{fig:w0wathawing} we show the $w_0w_a$ and $w_0w_a$-thawing posteriors for both of the BAO dataset combinations we analyze. They illustrate that a mild preference for $w_0 > -1$ persists even without the DESI data points that were claimed to drive the preference. For the DES-Y5 supernova dataset, the $95\%$ confidence interval does not include $\Lambda$CDM in either BAO dataset. We perform a principle components (PC) analysis indicated by the black line alone in the $w_0 \Omega_m$ degeneracy direction. Replacing the two DESI redshift bins in question with $z_{\rm eff} =0.51$ and $z_{\rm eff}= 0.706$ shifts the PC amplitude as shown in Fig.~\ref{fig:alpha}. For the BAO data to be compatible with $\Lambda$CDM (i.e. $w = -1$)  requires 
\begin{align}
\label{eq:A2}
\Omega_m &\approx 0.289 + \alpha, \,\,\, \,\,\,\, \text{($w_0w_a$)} \\
\Omega_m &\approx 0.299 + \beta, \,\,\,\, \,\,\,\, \text{($w_0w_a$-thawing)} 
\label{eq:A3}
\end{align}
where $\alpha$ and $\beta$ are the PC amplitudes, shown in Fig.~\ref{fig:alpha}. However, the supernova datasets have a strong preference for $\Omega_m > 0.29$, except for Pantheon, where $\Omega_m \approx 0.28$ is acceptable. This pull in opposite directions creates tension with $\Lambda$CDM that the two intermediate redshift DESI data points exacerbate. 

In Fig.~\ref{fig:tension2}, we repeat the analysis performed in Fig.~\ref{fig:tension} for the $\Lambda$CDM, $w_0w_a$, $w_0w_a$-thawing, and $w_0w_a$-tracking models. In the top plot of Fig.~\ref{fig:tension2}, we show the results of the $Q_{\rm{MAP}}$ goodness of fit test for each cosmological model and for each combination of data as quantified by the effective $n_{\sigma}$ values. In the $Q_{\rm{MAP}}$ test, we first find the overall best-fit point in the CMB + BAO + SN MCMC chain. This is referred to as the maximum a posteriori (MAP) point. We then use the $\chi^{2}_{\rm{SN}}$ value at this MAP point for the goodness of fit test. In addition, we calculate the effective number of degrees of freedom from the data by estimating the number of parameters constrained by the supernova data given that the CMB and BAO data already constrain the cosmological parameters. 

In general, we find qualitatively similar results to those results obtained in Fig.~\ref{fig:tension}. In particular, all of the models explored are consistent with the measured data and the variations across different cosmological models is smaller than the variations between data sets. This is caused by the relatively small difference in $\chi^2$ values, which are smaller than $\approx 10$, compared to the larger number of supernova data points, which are order 1000s for Pantheon, Pantheon+, and DES-Y5. The one exception is the $\Lambda$CDM point for Union3, which is slightly higher than the value for the evolving dark energy models. This difference arises because Union3 uses 22 distance bins instead of individual supernova light curves, so the improvements in $\chi^2$ for the evolving dark energy models are large enough to shift the $n_{\sigma}^{Q_{\rm{MAP}}}$ values.  

In the bottom panel of Fig.~\ref{fig:tension2}, we quantify the level of agreement between posterior distributions resulting from supernova alone to the posterior distributions from CMB + DESI. Because we are examining extended cosmological models and are using subsets of the data such as using only supernova data, the parameter posteriors are in general non-Gaussian. We, therefore, use Kernal Density Estimates (KDE) to account for this non-Gaussianity \cite{Raveri/etal:2021}. For all cases, we assess the consistency of the parameters $\Omega_m$ and $H_0$. For $w_0w_a$, we also include $w_0$ and $w_0+w_a$, while for the tracking and thawing cases, we include only the $w_{\rm{tracking}}$ or $w_{\rm{thawing}}$ parameters. 

For Pantheon+, DES-Y5, and Union3, the level of tension found when assuming the $\Lambda$CDM model is approximately $n_{\sigma}^{\rm{NG}}\approx 2-3\sigma$. As in the scalar field dark energy cases, the evolving dark energy models reduce this tension. $w_0w_a$-thawing is mostly preferred to $w_0w_a$-tracking.
For Pantheon+ and DES-Y5, the $w_0 w_a$-thawing model reduces the internal parameter tension slightly more than $w_0w_a$, though $w_0w_a$ has lower tension for Union3. For Pantheon, $\Lambda$CDM has the lowest level of tension across all of the models, another indicator for the source of tension being due to the shift in $\Omega_m$ in the newer supernova data set compilations. 

% --------------------------------------------------------------------
% --------------------------------------------------------------------
% --------------------------------------------------------------------
% --------------------------------------------------------------------
% --------------------------------------------------------------------
% --------------------------------------------------------------------
% --------------------------------------------------------------------
% --------------------------------------------------------------------
% --------------------------------------------------------------------

\appendix

\bibliographystyle{JHEP}
\bibliography{refs}

\end{document}